\documentclass[twocolumn,aps,prb,10pt,floatfix,superscriptaddress]{revtex4-2}
\usepackage{graphicx}
\usepackage[usenames,dvipsnames]{xcolor}
\usepackage[colorlinks=true,citecolor=blue,linkcolor=blue,urlcolor=blue,filecolor=blue,linktocpage=true]{hyperref}
\usepackage{textcomp}
\usepackage{stackengine}
\usepackage{listings}
\usepackage[makeroom]{cancel}
\usepackage[normalem]{ulem}        
\usepackage{array}
\usepackage{enumerate}             
\usepackage[shortlabels]{enumitem}  
\usepackage{verbatim}

\usepackage{amsmath}
\usepackage{amsfonts}
\usepackage{amssymb}
\usepackage{bbold}
\usepackage{relsize}
\usepackage{mathrsfs}
\usepackage{mathtools}
\usepackage{IEEEtrantools}          

\usepackage{orcidlink}

\usepackage[utf8x]{inputenc}
\usepackage{braket}
\usepackage{siunitx}
\usepackage{soul}
\usepackage{dcolumn} 
\usepackage{dsfont}  

\usepackage{txfonts}   
\usepackage{xr-hyper}

\newcommand{\tens}[1]{%
  \mathbin{\mathop{\otimes}\limits_{#1}}%
}

\def\bs#1{\boldsymbol{#1}}			
\def\unit{\mathbf{1}}				
\def\eps{\varepsilon}				
\def\mcH{\mathcal{H}}				
\def\mcT{\mathcal{T}}				
\def\mcP{\mathcal{P}}				
\def\imi{\mathrm{i}}				

\begin{document}
\title{Non-adiabatic corrections to chiral charge pumping in topological nodal semimetals}
\author{Matej Badin\,\orcidlink{0000-0001-7487-9802}}
\email{mbadin@sissa.it}    
\affiliation{SISSA\,–\,Scuola\,Internazionale\,Superiore\,di\,Studi\,Avanzati,\,Via\,Bonomea\,265,\,34136\,Trieste,\,Italy}
\affiliation{Department of Experimental Physics, Faculty of Mathematics, Physics and Informatics, Comenius University in Bratislava, Mlynsk\'{a} Dolina F2, 842 48 Bratislava, Slovakia}

\date{\today}

\begin{abstract}
Studying many-body versions of Landau-Zener-like problems of non-interacting electrons in the Slater formalism for several $\bs{k}\cdot\bs{p}$ models
representing Weyl and Dirac semimetals, we systematically include non-adiabatic corrections to a quantum limit of chiral charge pumping in these models. In this paper, we show that relative homotopy invariant
[Sun, \emph{et al.}, Phys. Rev. Lett. \textbf{121}, 106402 (2018)] and Euler class invariant [Bouhon, \emph{et al.}, Nat. Phys. \textbf{16}, 1137 (2020)] 
non-trivially manifest in the non-adiabatic corrections to the quantum limit of chiral charge pumping. 
These corrections could affect conductivity channels connected with the presence of chiral anomaly. 
Moreover, we show that for non-symmorphic systems this contribution is sensitive to the direction of the 
applied magnetic field (in respect to the so-called non-symmorphic nodal loop)
suggesting that the conjectured direction-selective chiral anomaly in non-symmorphic systems
[Bzdu\v{s}ek, \emph{et al.}, Nature (London) \textbf{538}, 75 (2016)] 
could lead to a strongly anisotropic longitudinal magnetoresistance. The presented approach can be easily applied to other $\bs{k}\cdot\bs{p}$ or tight-binding models.
\end{abstract}
\maketitle
\section{Introduction}
The characteristic feature of Weyl semimetals~\cite{Armitage2018,AnnualRev2018} (WSMs) is the chiral anomaly~\cite{Wan2011,Rylands2021},
experimentally manifested by a decrease of resistance in the presence of parallel electric and magnetic 
fields~\cite{Hosur2013,Zyuzin2012,Son2013,Ashby2013,Lu2015,Tabert2016,Tabert_Carbotte2016,Dai2017}.
In the quantum limit of strong magnetic fields [when only a chiral Landau level (LL) is occupied],
the chiral anomaly of WSM follows from the presence of chiral LLs
which connect the conduction bands to the valence bands~\cite{Deng2019,Deng2019_2,Das2020}. 
A pair of counter-propagating chiral Landau levels also appears
in Dirac semimetals~\cite{Armitage2018}, where a similar signature in magnetoresistance is expected if the relaxation time
of electrons is sufficiently large.  Curiously, it has been reported that a certain class of nodal-ring semimetals in non-symmorphic
systems~\cite{Bzdusek2016} should also exhibit chiral LLs, but only for special high-symmetry directions of the applied magnetic field.

In a WSM, in the adiabatic limit (for sufficiently weak electric fields), the motion of electrons can be modeled by the semiclassical 
approach~\cite{Hosur2013,Zyuzin2012,Son2013}, where the presence of a chiral LL implies pumping of electrons from occupied
to unoccupied bands at rate~\cite{Hosur2013,Zyuzin2012,Son2013}:
\begin{equation}
\frac{\partial Q}{\partial t} = -\chi V \frac{e^3}{4\pi^2\hbar^2} \bs{E} \cdot \bs{B}\,\text{,}\label{eq:pumping_rate}
\end{equation}
where $-e$ is the electronic charge, $\chi$ is the chirality of Weyl node ($\chi = \pm 1$), $V$ is the volume of the sample and $\bs{E}$
and $\bs{B}$ are the electric and magnetic fields, respectively. 
Due to the Nielsen-Ninomiya theorem~\cite{Nielsen1981}, the Weyl points (WPs) can appear only in pairs of
opposite chiralities. 
In the semiclassical Boltzmann formalism, Son and Spivak~\cite{Son2013} showed that the chiral charge pumping~[Eq.~(\ref{eq:pumping_rate})]
of two nodes possessing the opposite chiralities can be stabilized by an inter-node scattering of a finite relaxation time $\tau$ leading to
a finite field-dependent contribution to a conductivity~\cite{Son2013} 
\begin{equation}
\Delta\sigma(B) = \frac{e^3}{2\pi^2 \hbar} B\tau.
\end{equation}

In the quantum limit (when only the chiral LL is occupied)
and if the adiabatic approximation holds (for sufficiently weak electric fields),
this formula holds for every pair of Weyl nodes. Recently, the deviations from both
limits have been studied in Refs.~\cite{Deng2019,Deng2019_2,Das2020,Hwang2021,Das2022,Zeng2022}. 

Authors of several references~\cite{Deng2019,Deng2019_2,Das2020} examined the case when multiple LLs 
become occupied as the cyclotron frequency (magnetic field) is lowered compared
with the chemical potential (Fermi level) as one approaches a so-called classical regime (weak magnetic fields),
in which $\Delta\sigma(B) \propto B^2$. In this intermediate regime (between the classical and quantum one),
quantum oscillations, periodic in $1/B$, are present in longitudinal and planar transport
coefficients~\cite{Deng2019,Deng2019_2,Das2020}.
It has been proposed that they can act as a strong fingerprint of WSM~\cite{Deng2019}. 

Morever, Deng \emph{et al.}~\cite{Deng2019_2} showed
that longitudinal magnetoconductivity follows $\cos^6 \theta$
and $\cos^2 \theta$ dependencies in the weak and strong magnetic
field regimes, respectively. In Ref.~\cite{Deng2019_2} the
angle $\theta$ denotes the angle between the magnetic and electric fields
and thus the dependence deviates from the one predicted by Eq.~(\ref{eq:pumping_rate}).
Here, we focus purely to the non-adiabatic contributions to the chiral charge pumping
in the quantum limit and for the parallel case ($\bs{E} \parallel \bs{B}$). 

Das \emph{et al.}~\cite{Das2020} later
generalized the results of Refs.~\cite{Deng2019,Deng2019_2} to all magnetotransport coefficients
in this intermediate regime.

Hwang \emph{et al.}~\cite{Hwang2021} considered a situation when the electric field is arbitrarily
large, leading to a general Landau-Stark resonance. The authors of Ref.~\cite{Hwang2021} considered
both Landau and Stark quantization of energy levels for a minimal model of tight-binding WSM
and then studied nonequilibrium quantum transport on impurities using the Keldysh-Dyson
formalism. They showed that their minimal tight-binding model does not allow a Landau-Zener
transition between different LLs since the Berry connection is strictly zero (between different LLs). 
However, this does mean that a Landau-Zener transition is prohibited in a general model of a WSM
or Dirac semimetal.

Moreover, the usual semiclassical Boltzmann formalism or respectively the formula given by Eq.~(\ref{eq:pumping_rate}) becomes inapplicable 
if a Landau spectrum of a considered model shows occasional as well as symmetry-protected degeneracies
(when adiabatic approximation can fail). For example, this is exactly the case for the 
so-called \emph{non-symmorphic nodal loops} (NSNLs) and \emph{nodal chains} proposed in Ref.~\cite{Bzdusek2016}.
Therefore, a way that allows us to incorporate non-adiabatic corrections into the calculation of the chiral charge pumping
is sought. 

Recently, Das \emph{et al.}~\cite{Das2022} examined the case of second-order nonlinear magnetoconductivity
in type-I WSMs and multi-WSMs. Namely, they showed in the usual semi-classical Boltzmann formalism
for both considered classes of WSMs that there is no second-order chiral charge pumping in the electric field,
see eq.~(9) in Ref.~\cite{Das2022}, rising from a non-equilibrium distribution function used in the Boltzmann
transport formalism. Here, the second-order corrections to chiral charge pumping arise from the finite-time (non-adiabatic) contributions to
the time evolution of electron states which is modeled using the many-electron time-dependent Schrödinger equation (TDSE), see later.

Moreover, Zeng \emph{et al.}~\cite{Zeng2022} examined the case of nonlinear
longitudinal magnetoconductivity contributions rising from nonlinear planar effects
when $\bs{E} \nparallel \bs{B}$ but still $\bs{E}  \cdot \bs{B} \neq 0$,
e.g., a contribution to $\bs{j}_z$ of type $\chi_{zzx} E_z E_x$. Surprisingly, they find
it to be magnetic-field independent in the quantum limit, see eq. (19) in Ref.~\cite{Zeng2022},
and leading to two length scales in $1/B$-quantum oscillations.

This paper is organized as follows. In Sec.~\ref{sec:method}, we describe a simple numerical method that enables us to include
non-adiabatic corrections given by Eq.~(\ref{eq:pumping_rate}) (or its analogs). 
Namely, we show how to calculate the expectation value of the number of electrons in the
conduction bands still within the quantum limit but allowing for corrections implied by a finite
value of the electric field (still sufficiently weak).

In Secs.~\ref{sec:rhi} and~\ref{sec:eci}, we show that simple models (fulfilling
the Nielsen-Ninomiya theorem~\cite{Nielsen1981}; thus possessing two Weyl nodes
of the opposite chirality) that do and do not possess
the \textit{relative homotopy invariant}~\cite{Sun2018} or \textit{Euler class invariant}~\cite{Bouhon2020}
show a non-trivial signature in the mentioned non-adiabatic corrections to the chiral charge
pumping possibly measurable in experiments. Namely, we show that the deviations from
the expected values take place before the onset of annihilation of WPs (with respect
to the model free parameters). We note that we implicitly assume the intraband and internode
scattering times to be larger than the typical scale in which an electron passes through two Weyl nodes
(limit of sufficiently clean samples).

Finally, in Sec.~\ref{sec:nsnl} numerical results are presented for the class of Dirac semimetals possessing
so-called NSNLs~\cite{Bzdusek2016}. Curiously, a similar $1/B$ oscillation
pattern is observed in the chiral charge pumping compared with the quantum oscillations between the quantum and
classical limits in Refs.~\cite{Deng2019,Deng2019_2,Das2020},
even here of completely different origin (here purely due to a gap closing, there due to oscillations in
the density of states).

\section{Method \label{sec:method}}

We consider a simple many-body version of Landau-Zener-like problems of non-interacting electrons in $\bs{k}\cdot\bs{p}$
models under the presence of magnetic and electric fields. The first implies the presence of an infinite ladder of LLs,
the latter implies that we allow electrons to tunnel to neighboring LLs. We set the Fermi level to zero energy
$\varepsilon_{\mathrm{F}} = 0$ and consider the time-evolution of Slater determinants of momentum-dependent
Hamiltonians under the presence of an electric field. We consider only Slater determinants of electron states in which all
electrons share the same value of momentum $k_z$.

The presence of the electric field is modeled semi-classically as  
\begin{equation}
\partial_t k_z = -\frac{eE}{\hbar}\,\text{,}\label{eqn:momentum-increase}
\end{equation}
meaning that the electrons move along the one-dimensional bands (1D) opposite to the direction of $\mathbf{E}$. 
The semi-classical motion is immediately solved as 
$k_z(t) = -eEt/\hbar$. The presence of the magnetic field is modeled by the standard Peierls substitution in the corresponding
$\bs{k}\cdot\bs{p}$ model. The time evolution of the system is modeled in the quantum limit following the TDSE:
\begin{align}
    i\hbar\partial_t\ket{\psi(k_z)} &= \mathcal{H}^{\mathrm{many}}(k_z)\ket{\psi(k_z)}\,\text{,}\label{eq:ManySchR}
\end{align}
where $k_z(t) = -eEt/\hbar$, and $\ket{\psi(k_z)}$ is a many-electron wave function expressed in a basis of Slater determinants. 
The state $\ket{\psi(k_z)}$ is initially in the limit $t \to -\infty$ (replaced by a finite numerical value in numerical calculations)
set to the state in which all LLs corresponding to the valence bands are completely filled. 

Under the assumption of non-interacting electrons, the many-electron Hamiltonian $\mathcal{H}^{\mathrm{many}}(k_z)$ is the antisymmetric part
of a tensor product of the single particle Hamiltonians $\mathcal{H}(k_z)$. Hence, it follows that
the time evolution of the many-body state $\ket{\psi(k_z)}$ can be expressed using the time evolution of 
eigenstates of (single-electron) $\mathcal{H}(k_z)$ only. 

For every considered $\bs{k}\cdot\bs{p}$ model, we calculate the expectation value $N_{\mathrm{ex}}(E,B)$
of the number of electrons in the conduction bands in the limit $t \to \infty$ (replaced by a finite value in numerical
calculations). In the semiclassical limit, this expectation value enters the pumping rate through the relation:
\begin{align}
   \frac{\partial Q}{\partial t} = -\mathcal{D}\frac{N_{\mathrm{ex}}(E,B)e^2E}{\Delta k_z \hbar} =  -N_{\mathrm{ex}}(E,B)\frac{e^3V}{4\pi^2 \hbar^2}EB\,\text{,}
\end{align}
where $\mathcal{D} = BL_xL_y/(2\pi\hbar/e)$ is the degeneracy of LL,
$\Delta k_z = 2\pi / L_z$ is the spacing of the available states in the $k$-representation,
$V = L_x L_y L_z$ is the volume of a sample, and $L_x$, $L_y$ and $L_z$ are linear dimensions of the sample.

Note that any choice of  model under the presence of a magnetic field implies an infinite ladder of LLs below and above
the Fermi level. Therefore, we limit ourselves to only $N$ above and below it. The fermionic subspace of $\mathcal{H}^{\mathrm{many}}$ still has the dimensionality of $S = \binom{2N}{N} > 2^{N}$. To avoid this scaling, we express the many-electron wave function
in the basis of a limited number of Slater determinants created from the single-electron eigenstates
$\ket{\psi_{\mathrm{n}}\left(k_z\right)}$ of $\mathcal{H}(k_z)$. 

We limit only to the Slater determinants containing up to
$M$ $(M < N)$ electrons in the excited ($\eps_\textrm{n} > 0$) single-particle states.
That creates a basis of $S_M = \sum_{i=0}^{M}\binom{N}{i}^2 \sim N^{M}$
Slater determinants $\ket{\varphi_i\left({k_z}\right)}$.  The many-electron wave function
$\ket{\psi\left({k_z}\right)}$ is then expressed as
\begin{equation}
    \ket{\psi\left(k_z\right)} = \sum_i^{S_M} c_i(k_z)\ket{\varphi_i\left(k_z\right)}\,\text{,}
\end{equation}
where $\ket{\varphi_i\left(k_z\right)}$ is a Slater determinant of the form:
\begin{align}
    \ket{\varphi_i\left(k_z\right)} = \frac{1}{\sqrt{N!}}\sum_\tau \mathrm{sign\,\tau}
    \left\{\tens{j} \ket{\psi_{\mathrm{\tau(\alpha_i(j))}}\left(k_z\right)}\right\}\text{,} 
\end{align}
in which $\mathrm{sign\,\tau} = \pm 1$ is the parity of permutation $\tau$, and $\alpha_i(j)$ goes over
the set of $N$ one-electron eigenstates from which the Slater determinant is constructed. 

Under the assumption of non-interacting electrons, the Slater determinant $\ket{\varphi_i\left(k_z\right)}$ is an eigenstate
of $\mathcal{H}^{\mathrm{many}}(k_z)$ with energy $\varepsilon_i(k_z) = \sum_{j=1}^{N} \varepsilon_{\alpha(j)}(k_z)\text{.}$
It follows that the many-electron TDSE [Eq.~($\ref{eq:ManySchR}$)] (almost) diagonalizes in Slater formalism into the simple system of dimensionality $S_M$:
\begin{align}
    i\hbar C_{ij}^{-1}\partial_t c_j(k_z)= \varepsilon_i(k_z)c_i(k_z)\text{,}\label{eq:Master_TDSE}
\end{align}
where $C_{ij}$ is the overlap matrix between the eigenstates of $\mathcal{H}^{\mathrm{many}}(k_z+\Delta k)$ and
the eigenstates of $\mathcal{H}^{\mathrm{many}}(k_z)$. 

The solution of Eq.~($\ref{eq:Master_TDSE}$) only requires us to find a suitable time propagator $\mathcal{U}_{ij}(t+\Delta t,t)$
which under Eq.~($\ref{eqn:momentum-increase}$) can be written as $\mathcal{U}_{ij}(k_z-\Delta k, k_z)$.

The time evolution can then be expressed only as a problem of evolution of basis set coefficients
and a simple matrix-vector equation given, a suitable propagator $\mathcal{U}_{ij}(k_z-\Delta k, k_z)$ exists:
\begin{align}
    c_i(k_z-\Delta k) = \mathcal{U}_{ij}(k_z-\Delta k,k_z)c_j(k_z)\,\text{.}\label{eq:problem}
\end{align}
In this paper, we use a simple Euler forward propagator:
\begin{align}
 \mathcal{U}_{ij}(k_z-\Delta k,k_z) &= \sum_{n=0}^{S_M} \braket{\varphi_i(k_z-\Delta k)|\varphi_n(k_z-\frac{\Delta k}{2})}\nonumber\\
 &\times\exp{\left(\frac{i\Delta k_z}{\widetilde{E}}\varepsilon_n(k_z-\frac{\Delta k}{2})\right)}\nonumber\\
 &\times\braket{\varphi_n(k_z-\frac{\Delta k}{2})|\varphi_j(k_z)}\,\text{.}\label{eq:EulerPropagator}
\end{align}
and achieve the unitarity of the time evolution by the normalization of the wave function
after every step. 

Performing the real-time integration given by~Eqs.~(\ref{eq:problem})
and (\ref{eq:EulerPropagator}) we can find the expectation value $N_{\mathrm{ex}}(E,B)$
in the limit $t \to \infty$.

Since the dot product of two Slater determinants can be calculated in $\mathcal{O}(N^3)$ steps, see Appendix~\ref{sec:dot_product_Slater},
the evolution can be evaluated in $\mathcal{O}(S_M^2 N^3) \sim \mathcal{O}(N^{2M+3})$ steps. 

The results of the numerical algorithm are verified on a single-node isotropic WP $\bs{k}\cdot\bs{p}$ model
presented in Appendix~\ref{sec:demonstation_WSM}.

\section{Relative homotopy invariant \label{sec:rhi}}
\subsection{Models}

In the presence of a mirror symmetry $m_z: (x,y,z) \mapsto (x,y,-z)$, a pair of mirror-related WPs carry opposite chirality. 
While it is generally expected that such WPs pairwise annihilate at the symmetric plane upon collision, it has been reported~\cite{Sun2018}
that a finer relative homotopy invariant may prevent their annihilation, enforcing instead their conversion into a nodal loop (NL).
Such a scenario prominently arises for two-band models based on a pair of orbitals with different $m_z$ eigenvalue~\cite{Lim2017},
while a trivial annihilation occurs if the two orbitals have the same mirror eigenvalue.

We consider minimal $\bs{k}\cdot\bs{p}$ models belonging to both topological classes, assuming for simplicity
an additional $\mathsf{SO}(2)$ symmetry around the mirror normal. For the trivial case with mirror operator $\hat{m}_z = \mathds{1}$, we take
\begin{equation}
\mcH_A(\bs{k}) = \hbar v \left(k_x \sigma_x + k_y \sigma_y\right) + (m - \alpha k_z^2)\sigma_z, \label{eq:Ham_A}
\end{equation}
which for $m/\alpha>0$ exhibits a pair of WPs at $(0,0,\pm\sqrt{m/\alpha})$ that annihilate for $m=0$ at $\bs{k} = \bs{0}$.
For the non-trivial case with mirror operator $\hat{m}_z = \sigma_z$, the minimal model must include higher-order terms. We specifically take
\begin{align}
\mcH_B(\bs{k}) &= \alpha k_z (k_x \sigma_x + k_y \sigma_y ) \nonumber\\
 &+ \left(\beta\left(k_x^2 + k_y^2 - k_z^2\right) + m\right)\sigma_z, \label{eq:Ham_B}
\end{align}
which [like $\mcH_A(\bs{k})$] for $m>0$ exhibits a pair of WPs at $(0,0,\pm\sqrt{m/\beta})$, that [as opposed
to $\mcH_A(\bs{k})$] convert to a NL in $k_z = 0$ with radius $\kappa = \sqrt{-m/\beta}$ for $m/\beta<0$.
A two-band model like $\mcH_B(\bs{k})$ (but with twice the value of the relative homotopy invariant,
and with a flipped sign of the $k_z^2 \sigma_z$ term) has been considered for ferromagnetic HgCr$_2$Se$_4$~\cite{Xu2011}.
A recent work~\cite{Nelson2021} showed that the relative homotopy invariant of the band nodes in HgCr$_2$Se$_4$ is of a delicate topological character.

A slightly adjusted version of $\mcH_A(\bs{k})$ also governs the conversion of WPs into a NL in ZrTe,
where the parameter $m$ is manipulated by strain~\cite{Bouhon2020}; here,
the topological stability is enhanced by a $C_{2}\mcT$-protected non-Abelian topological invariant~\cite{Wu2019}.

\begin{figure}[t!]
	\includegraphics[width=9cm]{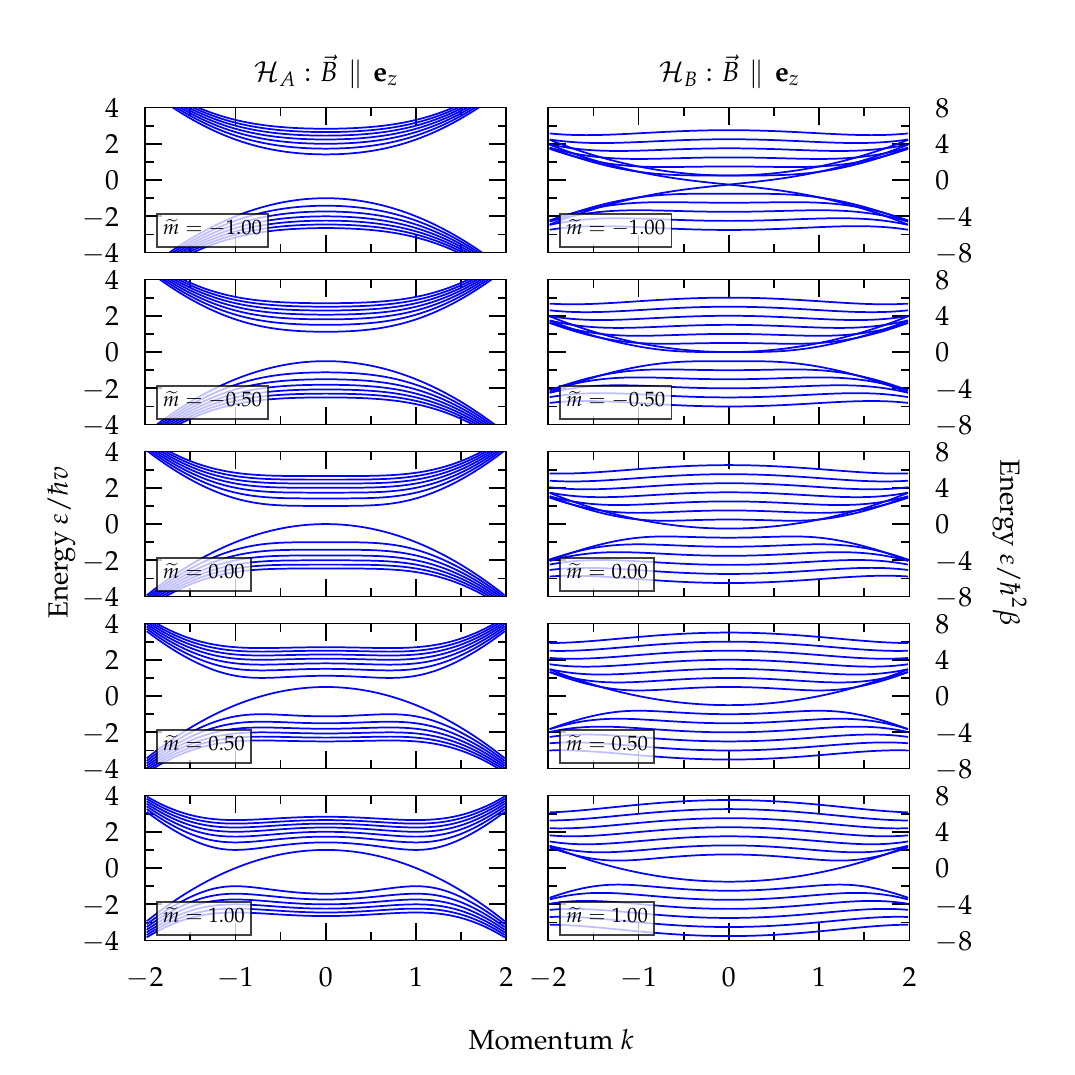}
	\caption{Landau Levels (LLs) of $\mcH_A(\bs{k})$ and $\mcH_B(\bs{k})$ for the magnetic field oriented along the z-axis and $\widetilde{B} = 0.5$ for various choices of $\widetilde{m}$ [$\widetilde{m} = m/\hbar v$ for $\mcH_A(\bs{k})$ and $\widetilde{m} = m/\hbar^2 \beta$ for $\mcH_B(\bs{k})$]. Note the pair of Weyl poits (WPs) for $\widetilde{m} > 0$ for $\mcH_A(\bs{k})$ and the manifestation of the nodal loop (NL) as a point in this orientation of magnetic field for $\mcH_B(\bs{k})$ and $\widetilde{m} < 0$. Seven LLs below and above the Fermi level are shown. \label{fig:LLs_AB_Bz}}
\end{figure}

\begin{figure}[t!]
	\includegraphics[width=9cm]{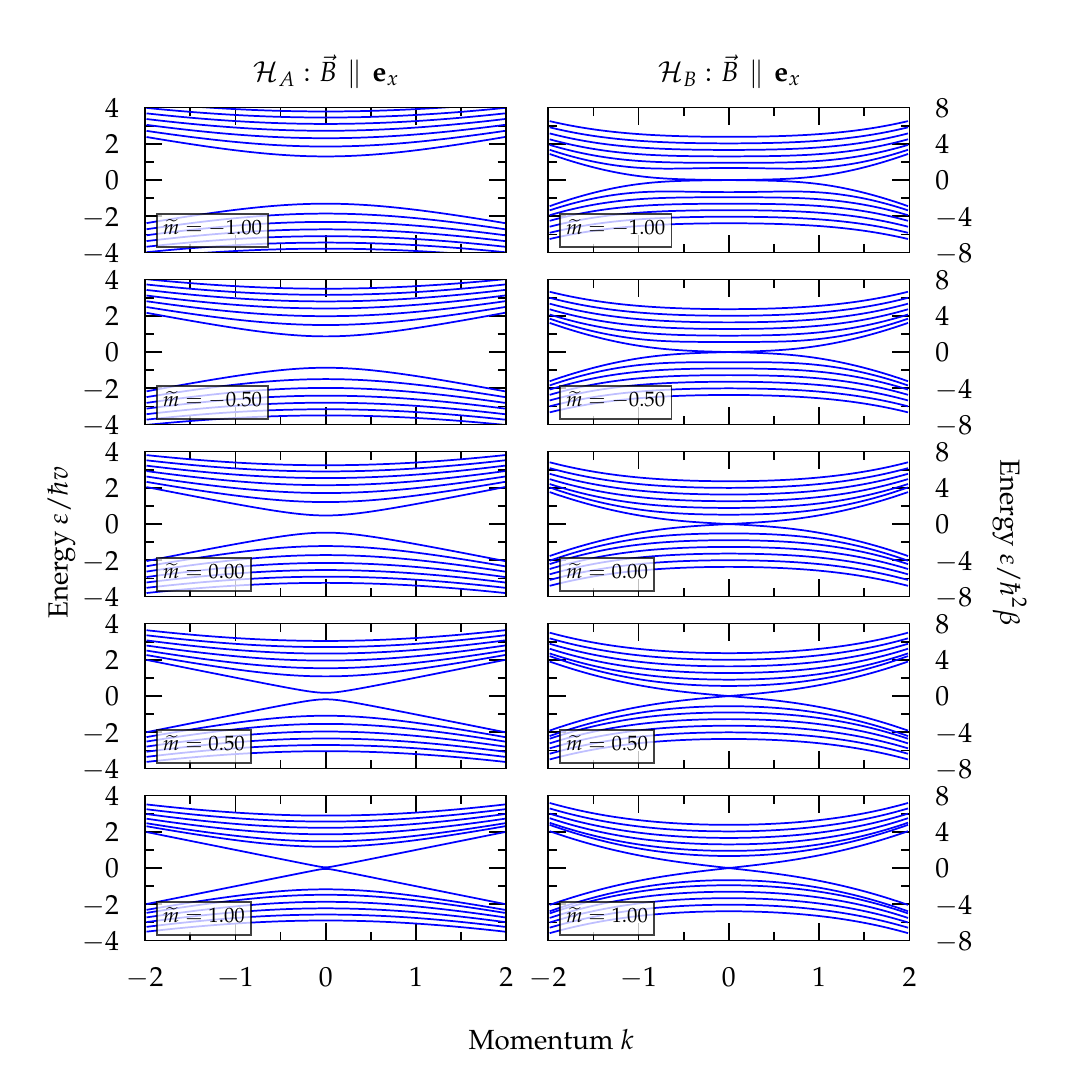}
	\caption{Landau Levels (LLs) of $\mcH_A(\bs{k})$ and $\mcH_B(\bs{k})$ for the magnetic field oriented along the x-axis and $\widetilde{B} = 0.5$ for various choices of $\widetilde{m}$ [$\widetilde{m} = m/\hbar v$ for $\mcH_A(\bs{k})$ and $\widetilde{m} = m/\hbar^2 \beta$ for $\mcH_B(\bs{k})$]. Note the nodal loop (NL) for $\widetilde{m} < 0$ and $\mcH_B(\bs{k})$. Seven LLs below and above the Fermi level are shown.\label{fig:LLs_AB_Bx}}
\end{figure}

LLs of the models are presented in Figs.~\ref{fig:LLs_AB_Bz} and \ref{fig:LLs_AB_Bx}.
In the numerical calculation we set the values of $\hbar v = \alpha = 1$ for $\mcH_A(\bs{k})$ and $\alpha = \hbar^2 \beta = 1$ for $\mcH_B(\bs{k})$.

\subsection{Pumping rate}
\begin{figure*}[t!]
	\includegraphics[width=0.99\textwidth,keepaspectratio]{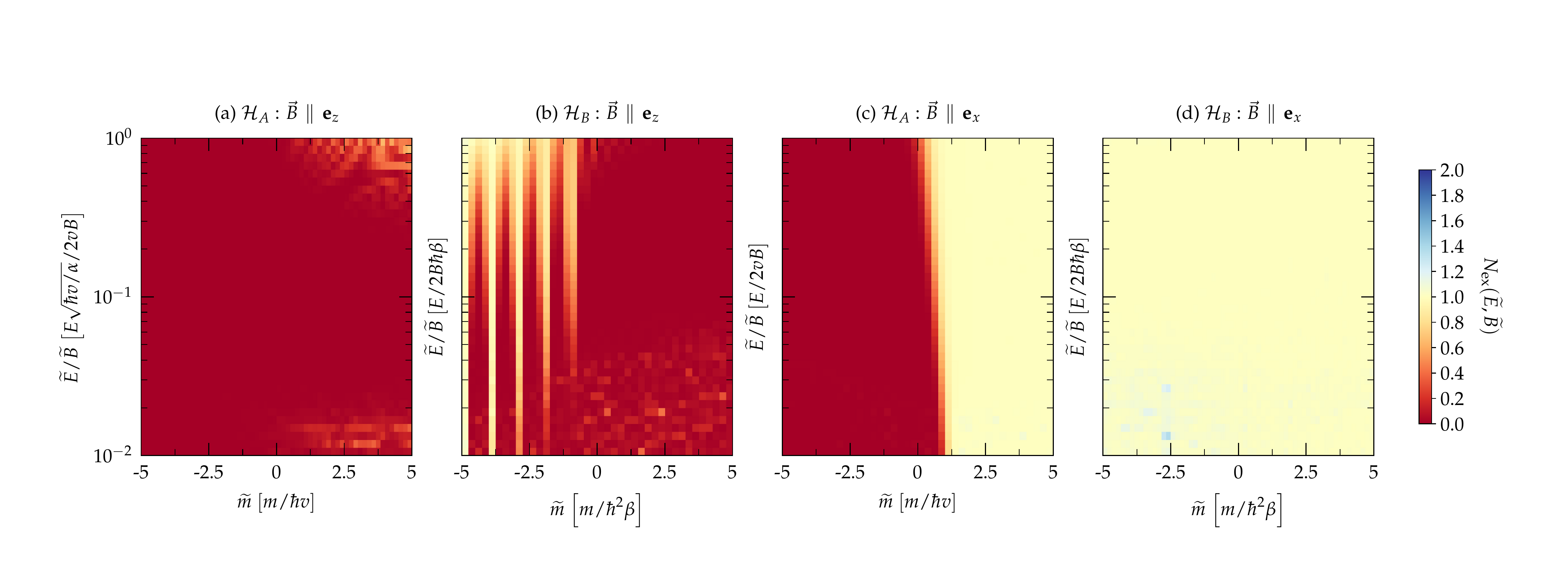}
	\caption{(a) - (d) The expectation value $N_{\mathrm{ex}}(\widetilde{E},\widetilde{B})$ of the number of electrons
	in the conduction bands as a function of the ratio of rescaled magnetic and electric fields $\widetilde{E}/\widetilde{B}$
	and the rescaled parameter $\widetilde{m}$ for the various orientations of magnetic field.  \label{fig:Ham_AB_phase_diagram}}
\end{figure*}

In Fig.~\ref{fig:Ham_AB_phase_diagram} we show the expectation value of excited electrons in the conduction bands
- $N_{\mathrm{ex}}(E, B)$ [or to be more precise - $N_{\mathrm{ex}}(\widetilde{E}, \widetilde{B})$, where
$\widetilde{E}$ and $\widetilde{B}$ are rescaled fluxes of electric and magnetic fields $E$ and $B$ (to be defined
later), respectively] - modifying the quantum limit of charge pumping created by the presence of chiral anomaly. 

While the model $\mcH_A(\bs{k})$ without the additional $\mathsf{SO}(2)$ symmetry around the mirror normal
does not show any significant additional structure in $N_{\mathrm{ex}}(\widetilde{E},\widetilde{B})$, see Fig.~\ref{fig:LLs_AB_Bz}(a),
the model $\mcH_B(\bs{k})$ possessing nontrivial $\mathsf{SO}(2)$ symmetry shows an additional structure for $\vec{B}\,\mathbin{\|}\,\mathbf{e}_z$ 
and $m < 0$, see the stripes in Fig.~\ref{fig:Ham_AB_phase_diagram}(b). This structure is caused
by gap closings at specific values of $\widetilde{m}$, e.g., $\widetilde{m} = -1$, see Fig.~\ref{fig:LLs_AB_Bz}. 
These periodic-in-$\widetilde{m}$ gap closings are related to the gapless band structure of the NL phase
and are protected by the mirror symmetry.

However, if one is given a WSM phase with two WPs that approach one another after tuning some parameter (e.g., $\widetilde{m}$),
then the $\widetilde{m}$-dependence of the pumping rate in applied $\vec{B}\,\mathbin{\|}\,\mathbf{e}_z$ does not reveal whether
the WPs would annihilate or convert into the NL semimetal: we find that the dependence on the WSM side of the phase diagram
[of $N_{\mathrm{ex}}(\widetilde{E}, \widetilde{B})$] is qualitatively the same for both $\mcH_A(\bs{k})$ and $\mcH_B(\bs{k})$, 
except the occasional gap closing at specific values of $\widetilde{m}$).

The models can be easily distinguished if $\vec{B}\,\mathbin{\|}\,\mathbf{e}_x$, where both models show different
structures of $N_{\mathrm{ex}}(\widetilde{E},\widetilde{B})$ with respect to the parameter $\widetilde{m}$, see Fig.~\ref{fig:Ham_AB_phase_diagram}(c) and 3(d).

Namely, observe that the pumping rate for the Hamiltonian $\mcH_B(\bs{k})$ with a non-trivial relative homotopy invariant is de facto constant
as a function of $\widetilde{m}$. In contrast, for $\mcH_A(\bs{k})$, where the WPs can annihilate,
we observe a decrease in pumping rate before
the WPs actually annihilate. This early vanishing of the chiral anomaly is particularly pronounced for small values of $\widetilde{E}/\widetilde{B}$
(i.e., weak electric field or strong magnetic field). This can be easily explained by the occurrence of a degenerate WP for $\mcH_A(\bs{k})$
while for $\mcH_B(\bs{k})$ a NL or a WP is always present with respect to $\widetilde{m}$, see Fig.~\ref{fig:LLs_AB_Bx}.

Furthermore, for sufficiently high values of $\widetilde{m}$, $\mcH_A(\bs{k})$ and $\vec{B}\,\mathbin{\|}\,\mathbf{e}_z$,
an interference pattern can be observed on $N_{\mathrm{ex}}(\widetilde{E},\widetilde{B})$, see Fig.~\ref{fig:Ham_AB_phase_diagram}(a).
It corresponds to the presence of the top-most valence band close to the conduction bands at $k = 0$, see Fig.~\ref{fig:LLs_AB_Bz} - the region around
$\widetilde{m} = 1$. The precise value of $\widetilde{m}$ for which such a constructive/destructive interference
appears is determined by the time the electron in the top-most valence band spends in vicinity of the conduction bands.

For the calculation of $N_{\mathrm{ex}}(\widetilde{E}, \widetilde{B})$, see Fig.~\ref{fig:Ham_AB_phase_diagram},
in each of two bands 7 LLs were considered
leading to an overall 13 LLs (after removing a ghost state). Slater determinants with up to $M = 2$ excited electrons
(implying overall 469 Slater determinants) were used and overall 1402 $k$-points were used in the simulation
of the real-time dynamics. Note that, for the proper identification of ghost states, the matrix representations of
$\mcH_A(\bs{k})$ and $\mcH_B(\bs{k})$ are truncated at a much higher number of states (80) than just
the number of considered LLs (13).

\section{Euler class invariant \label{sec:eci}} 
\subsection{Models}

It is known that $C_{2z} \mcT$ symmetry (composition of time reversal with $\pi$ rotation around the $z$ axis)
can stabilize WPs inside a symmetric plane $k_z = 0$ (or $k_z = \pi$), as  observed e.g.~in 
the $k_z \!=\! 0$ plane of $\textrm{WTe}_2$~\cite{Soluyanov2015}, $\textrm{MoP}$~\cite{Lv2017} and $\textrm{TaAs}$~\cite{Weng2015,Lv2015,Xu2015}.
In addition to their chiral charge, such WPs are also characterized by their Euler class invariant~\cite{Bouhon2020}.
A non-trivial value of the Euler class prevents pairwise annihilation of colliding WPs with the same chirality inside the symmetric plane.

We consider minimal $C_{2z}\mcT$-symmetric $\bs{k}\cdot\bs{p}$ models that describe a pair of colliding WPs
with opposite chirality with the trivial vs.~non-trivial Euler class. For the trivial case, we consider
\begin{align}
\mcH_C(\bs{k}) &= 2(\alpha k_x k_y - m) \sigma_x  \nonumber\\
&+ \hbar v \left[k_z \sigma_y + (k_x - k_y) \sigma_z\right],\label{eq:Ham_C}
\end{align}
which for $m/\alpha>0$ exhibits a pair of WPs of opposite chirality at $\pm(\sqrt{m/\alpha},\sqrt{m/\alpha},0)$
that annihilate for $m=0$ at $\bs{k} = \bs{0}$. We contrast this to a model with a non-trivial value of the Euler class, namely,
\begin{align}
\mcH_D(\bs{k}) &= 2(\alpha k_x k_y - m) \sigma_x + \beta k_x k_z \sigma_y \nonumber\\
&+ \gamma(k_x ^2 -k_y^2 - k_z^2) \sigma_z,\label{eq:Ham_D}
\end{align}
which [like $\mcH_C(\bs{k})$] for $m/\alpha>0$ exhibits a pair of WPs of opposite chirality at $\pm(\sqrt{m/\alpha},\sqrt{m/\alpha},0)$,
that [as opposed to $\mcH_C(\bs{k})$] collide at $\bs{k}=\bs{0}$ and bounce to form WPs at $\pm(\sqrt{-m/\alpha},-\sqrt{-m/\alpha},0)$ for $m<0$.
While we are unaware of material examples exhibiting Weyl points that scatter after collision due to non-trivial of the Euler class,
they could potentially be implemented and probed in cold-atom setups~\cite{Nal2020}.

\begin{figure}[t!]
	\includegraphics[width=9cm]{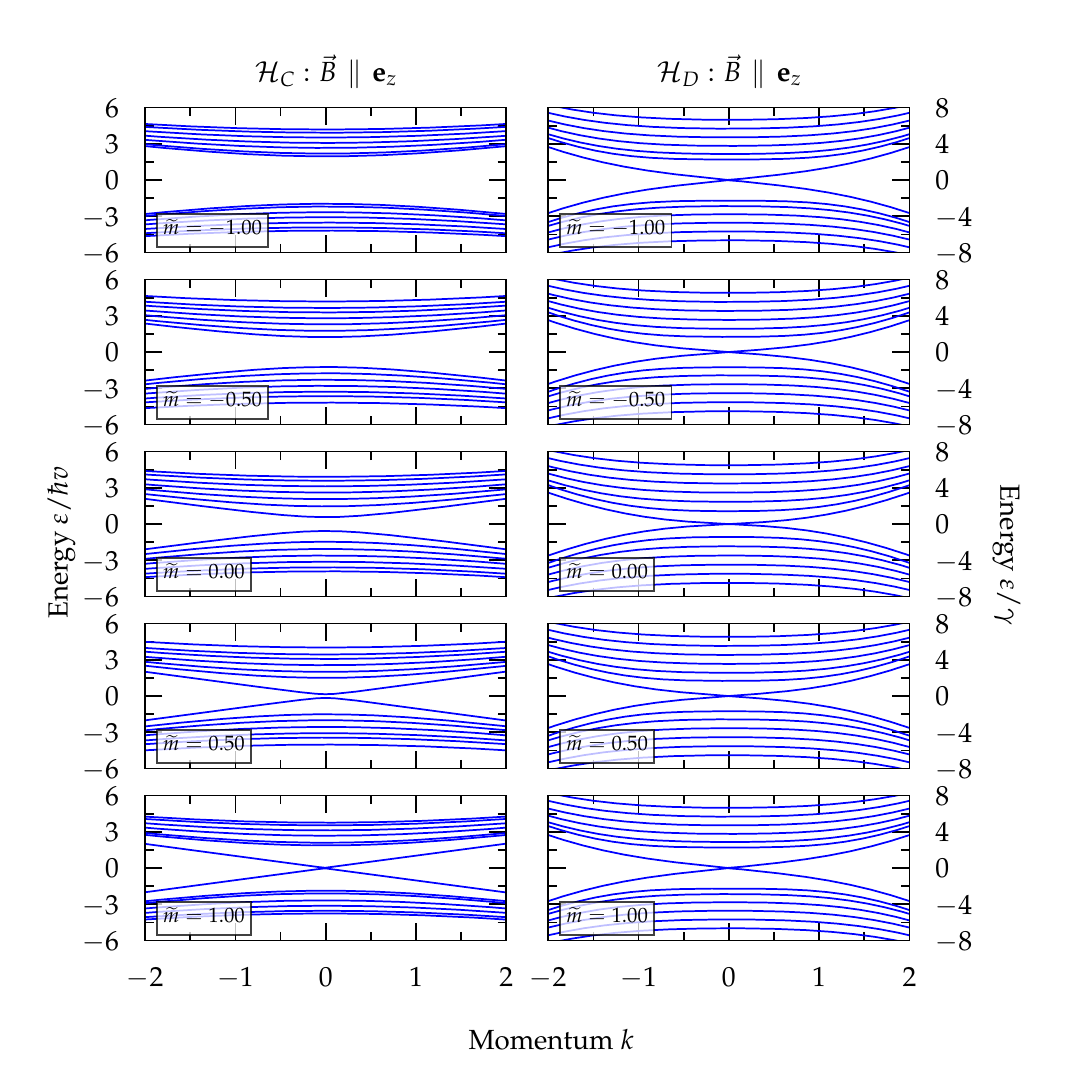}
	\caption{Landau Levels (LLs) of $\mcH_C(\bs{k})$ and $\mcH_D(\bs{k})$ for the magnetic field oriented along the z-axis and $\widetilde{B} = 0.5$ for various choices of $\widetilde{m}$ [$\widetilde{m} = m/\hbar v$ for $\mcH_C[\bs{k})$ and $\widetilde{m} = m/\gamma$ for $\mcH_D(\bs{k})$]. Seven LLs below and above the Fermi level are shown. \label{fig:LLs_CD_Bz}}
\end{figure}

\begin{figure}[t!]
	\includegraphics[width=9cm]{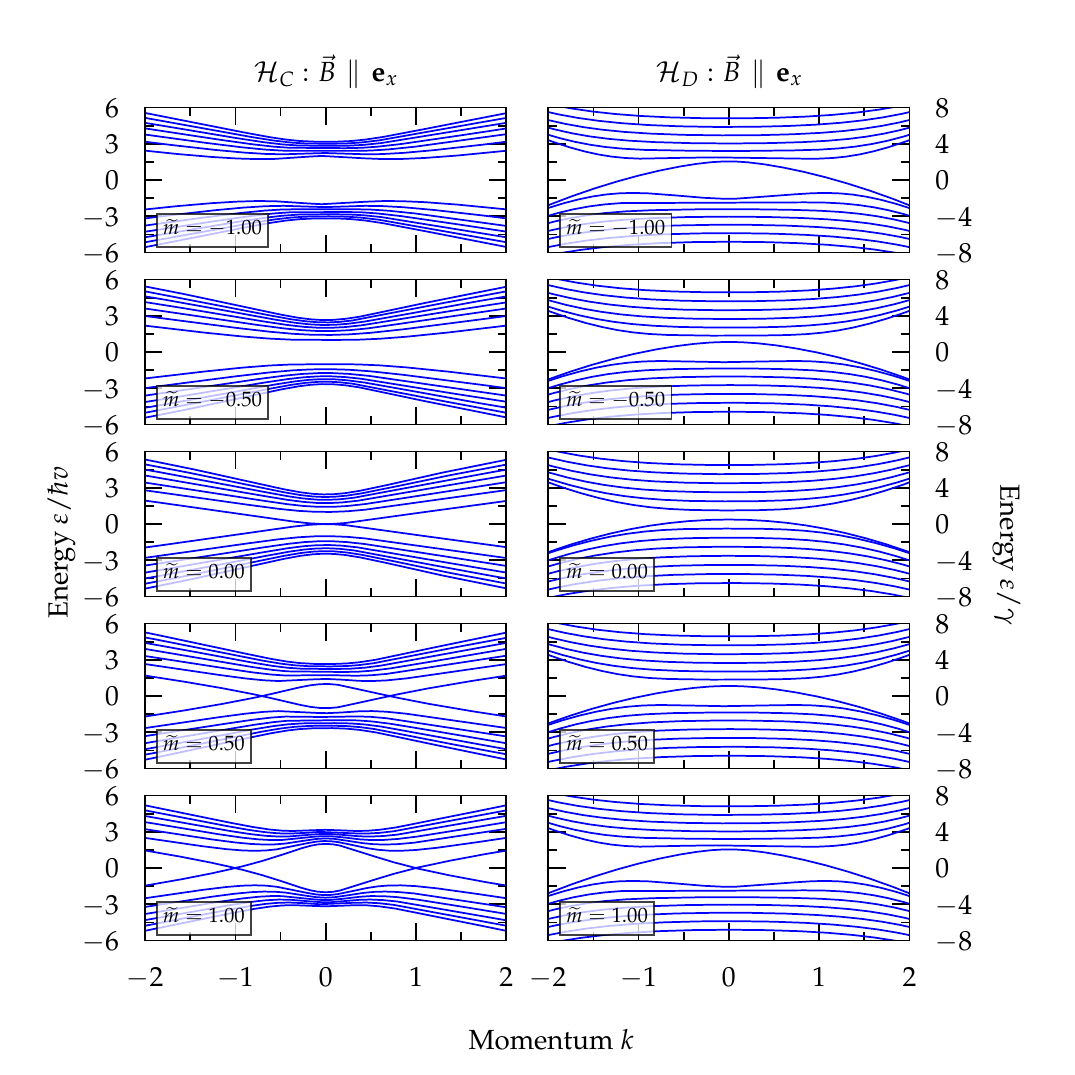}
	\caption{Landau Levels (LLs) of $\mcH_C(\bs{k})$ and $\mcH_D(\bs{k})$ for the magnetic field oriented along the x-axis and $\widetilde{B} = 0.5$ for various choices of $\widetilde{m}$ [$\widetilde{m} = m/\hbar v$ for $\mcH_C(\bs{k})$ and $\widetilde{m} = m/\gamma$ for $\mcH_D(\bs{k})$]. Seven LLs below and above the Fermi level are shown. \label{fig:LLs_CD_Bx}}
\end{figure}

LLs of the models are presented in Figs.~\ref{fig:LLs_CD_Bz} and \ref{fig:LLs_CD_Bx}.
In the numerical calculation we set the values of $\hbar v = \alpha = 1$ for $\mcH_C(\bs{k})$ and $\alpha = \beta = \gamma = 1$ for $\mcH_D(\bs{k})$.

\subsection{Pumping rate}

\begin{figure*}[t!]
	\includegraphics[width=0.99\textwidth,keepaspectratio]{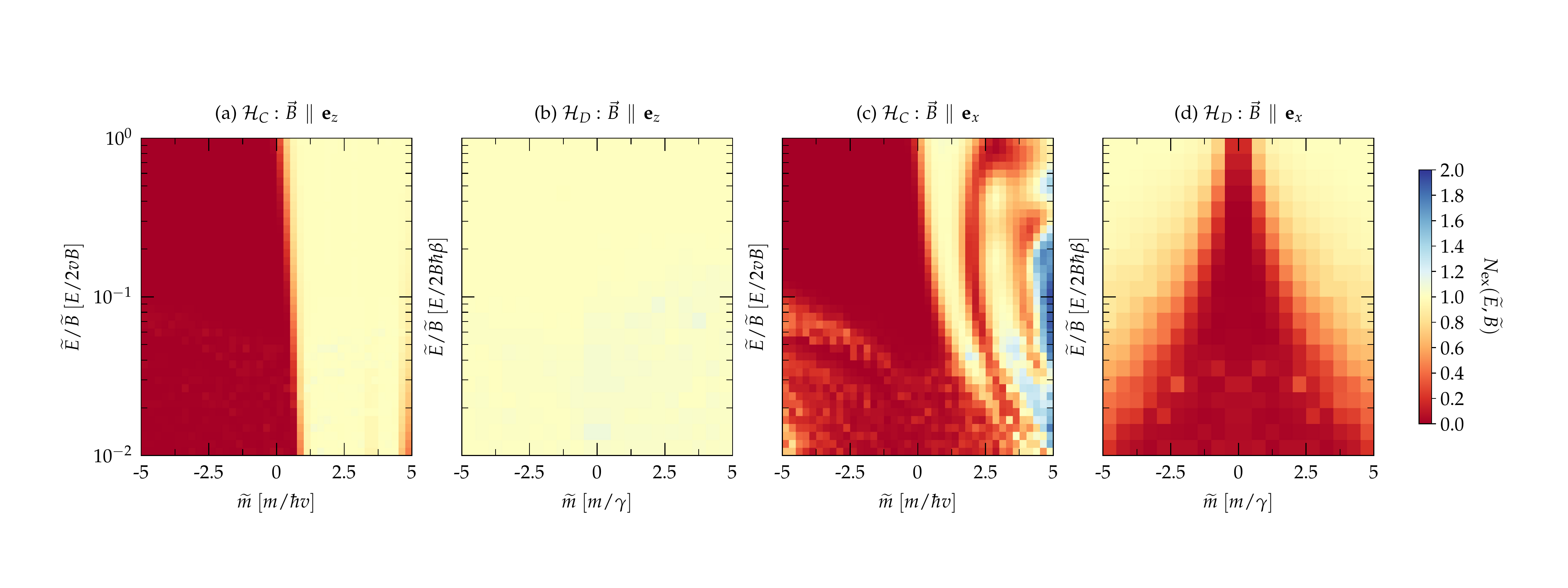}
	\caption{(a) - (d) The expectation value $N_{\mathrm{ex}}(\widetilde{E},\widetilde{B})$ of number of electrons in conduction bands as a function of the ratio of rescaled magnetic and electric field $\widetilde{E}/\widetilde{B}$ and the rescaled parameter $\widetilde{m}$ for various orientations of magnetic field. \label{fig:Ham_CD_phase_diagram}}
\end{figure*}

In Fig.~\ref{fig:Ham_CD_phase_diagram} we show the expectation value of excited electrons in the conduction bands - $N_{\mathrm{ex}}(\widetilde{E},\widetilde{B})$ - modifying the chiral charge pumping for models of trivial [$\mcH_C(\bs{k})$] and nontrivial [$\mcH_D(\bs{k})$]
Euler class. 

The models can be easily distinguished by the application of the magnetic field along any direction. If $\vec{B}\,\mathbin{\|}\,\mathbf{e}_z$, 
the nontrivial model [$\mcH_D(\bs{k})$] does not show any deviation from the value $N_{\mathrm{ex}}(\widetilde{E},\widetilde{B}) = 1$ while
the trivial one [$\mcH_C(\bs{k})$] differs in respect to the sign of $\widetilde{m}$. This behavior can be easily explained via the corresponding LLs.

For the trivial model, the deviation from the value $N_{\mathrm{ex}}(\widetilde{E},\widetilde{B}) = 1$ is caused by the gap opening (annihilation of a WP),
see Fig.~\ref{fig:LLs_CD_Bz}, as one decreases $\widetilde{m}$. On the contrary for non-trivial model, the WP is presented for any value of 
$\widetilde{m}$, see Fig.~\ref{fig:LLs_CD_Bz}. Namely, observe that, for the pumping rate for the Hamiltonian $\mcH_C(\bs{k})$ with the trivial Euler class, the 
hybridization of chiral LLs upon the change of parameter $\widetilde{m}$ is pronounced for small values of $\widetilde{E}/\widetilde{B}$ 
(i.e., weak electric field, strong magnetic field) before the WPs annihilate. Therefore, a decline of 
$N_{ex}(\widetilde{E}, \widetilde{B})$ from the expected limit - 1 (in here chosen units) suggests trivial class.

More interestingly, the models show nontrivial behavior if $\vec{B}\,\mathbin{\|}\,\mathbf{e}_x$. 
The spectrum of the nontrivial model [$\mcH_D(\bs{k})$]
is symmetric with respect to $\widetilde{m}$, see Fig.~\ref{fig:LLs_CD_Bx}, which manifests itself in the $N_{\mathrm{ex}}(\widetilde{E},\widetilde{B})$
as the function of $\widetilde{m}$, see Fig.~\ref{fig:Ham_CD_phase_diagram}(d). Note that a cusp visible at Fig.~\ref{fig:Ham_CD_phase_diagram}(d) disappears
for the ratio $\widetilde{E}/\widetilde{B} > 1$, if one overcomes the critical value of $\widetilde{E}$ which enables tunneling also for $\widetilde{m} = 0$.
The phase diagram of $N_{\mathrm{ex}}(\widetilde{E},\widetilde{B})$ for $\mcH_C(\bs{k})$, see Fig.~\ref{fig:Ham_CD_phase_diagram}(c), reveals a
non-trivial structure - the oscillations of $N_{\mathrm{ex}}(\widetilde{E},\widetilde{B})$ with respect to 
$\widetilde{m}$ which are caused by an interference.
The precise value of $\widetilde{m}$ for such constructive/destructive interference to happen is determined by the time the electron in
the top-most valence band spends in the vicinity of the conduction bands, see LLs in Fig.~\ref{fig:LLs_CD_Bx}.

For the calculation of $N_{\mathrm{ex}}(\widetilde{E}, \widetilde{B})$, see Fig.~\ref{fig:Ham_CD_phase_diagram}
the simulation parameters were the same as for the production of Fig.~\ref{fig:Ham_AB_phase_diagram}.

\section{NSNLs \label{sec:nsnl}}

\subsection{Model}
In this section, we revisit the chiral LLs levels of NLs protected on the boundary of the Brillouin zone
by a glide symmetry~\cite{Bzdusek2016}. The elementary model of a NSNL is 
\begin{equation}
\mcH_\textrm{NSNL}(\bs{k}) = \hbar v \left( k_x \Gamma_1 + k_y \Gamma_2 + k_z \Gamma_3\right) + w\Gamma_{34} \label{eqn:NSNL-Ham}
\end{equation}
where $\{\Gamma\}_{i=1}^5$ are pairwise anticommuting Dirac matrices squaring to $+\unit$, and $\Gamma_{ij} = -\tfrac{\imi}{2}[\Gamma_i,\Gamma_j]$.
NSNLs can be obtained from a parent Dirac node in a $\mcP\mcT$-symmetric crystal with screw rotation symmetry~\cite{Young2012,Yang2014b}
upon breaking the inversion symmetry. It has been argued by a theoretical work on these systems~\cite{Bzdusek2016}
that NSNLs exhibit LL crossing at zero energy for magnetic fields $\bs{B}$ applied within the glide plane.
Nevertheless, the chiral LLs are gapped if $\bs{B}$ deviates from the high-symmetry plane.
It has therefore been suggested that these models should exhibit a direction-selective chiral anomaly~\cite{Bzdusek2016}.

Here, we use the method presented in Sec.~\ref{sec:method} to investigate the disappearance
of the chiral-anomalous pumping rate of the Hamiltonian in Eq.~(\ref{eqn:NSNL-Ham}).

\begin{figure}[t!]
	\includegraphics[width=8.6cm]{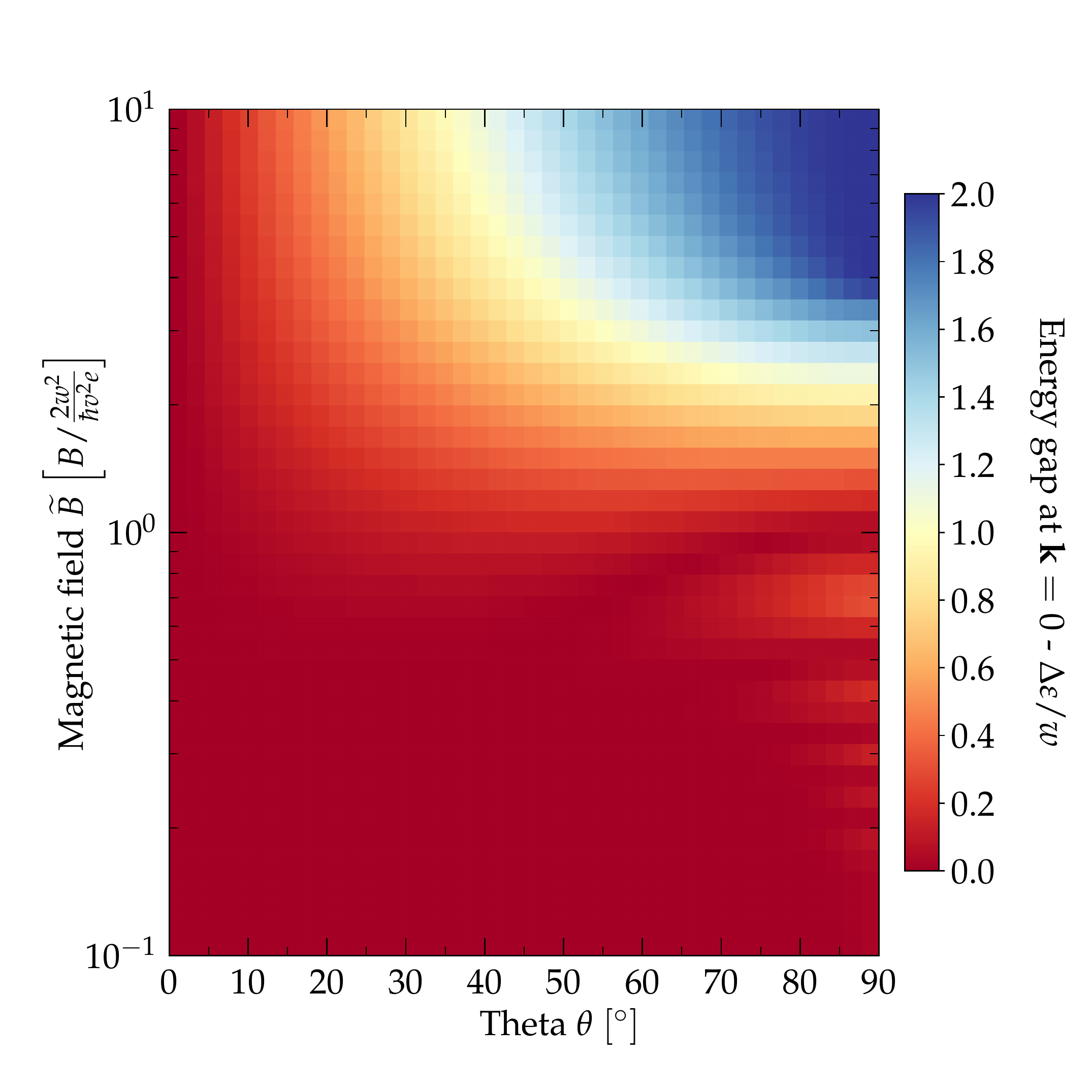}
	\caption{Energy gap $\Delta\varepsilon/w$ of $\mcH_\textrm{NSNL}(\bs{k})$ at $\bs{k} = 0$ as a function of the magnitude of the applied magnetic field $\widetilde{B}$ and the angle $\theta$ between the direction of the applied magnetic field and the non-symmorphic nodal loop (NSNL). Note the ladder of gap closing at $\theta = 90^{\circ}$ and $\widetilde{B} = 1/n$ for $n \in \mathbb{N}$; which manifests in $N_{\mathrm{ex}}(\widetilde{E},\widetilde{B})$, see also Fig.~\ref{fig:HSNL_phase_diagram_details}. \label{fig:HSNL_Gaps}}
\end{figure}

In Fig.~\ref{fig:HSNL_Gaps}, we show the band gap in LLs for $\mathbf{k} = 0$ as a function
of applied magnetic field and the angle $\theta$ between the direction of the applied magnetic field and the plane of NSNL in $k$-space.
LLs of the model as a function of the angle $\theta$ between the direction of the applied magnetic field and the plane of NSNL
in $k$-space are presented in Appendix ~\ref{sec:spectra_HSNL}.

\subsection{Pumping rate}
\begin{figure*}[t!]
	\includegraphics[width=0.99\textwidth,keepaspectratio]{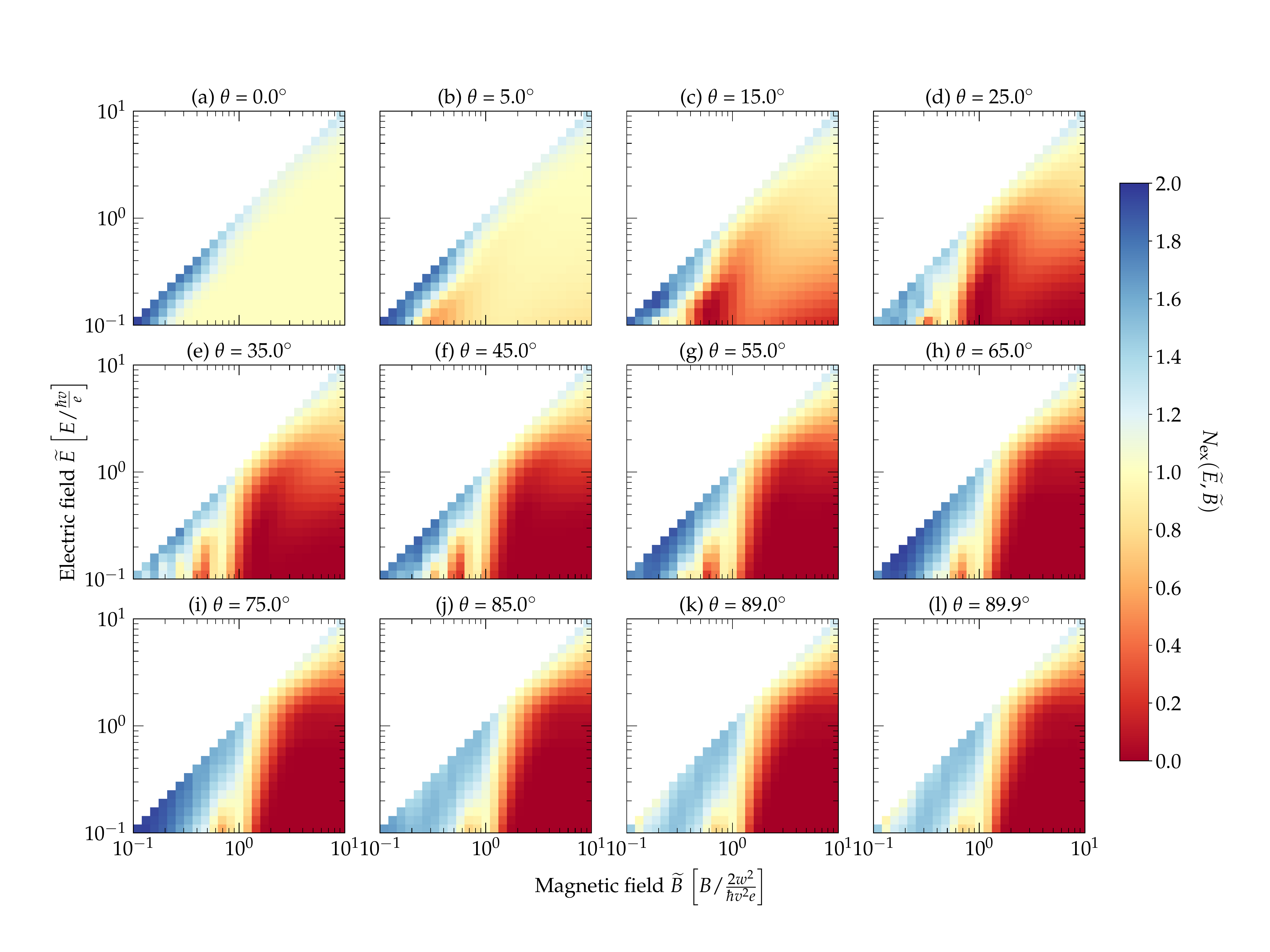}
	\caption{The expectation value $N_{\mathrm{ex}}(\widetilde{E},\widetilde{B})$ of the number of electrons in the conduction bands as a function of rescaled magnetic and electric fields $\widetilde{B}$ and $\widetilde{E}$. (a) - (l) Disappearance of chiral-anomalous pumping if the direction of the magnetic field deviates from the plane of non-symmorphic nodal loop (NSNL) in the region $\widetilde{B} > 1$. In~Fig.~\ref{fig:HSNL_phase_diagram_details}, we present details of phase diagrams for $\widetilde{B} < 1$. \label{fig:HSNL_phase_diagram}}
\end{figure*}

\begin{figure*}[t!]
	\includegraphics[width=0.99\textwidth,keepaspectratio]{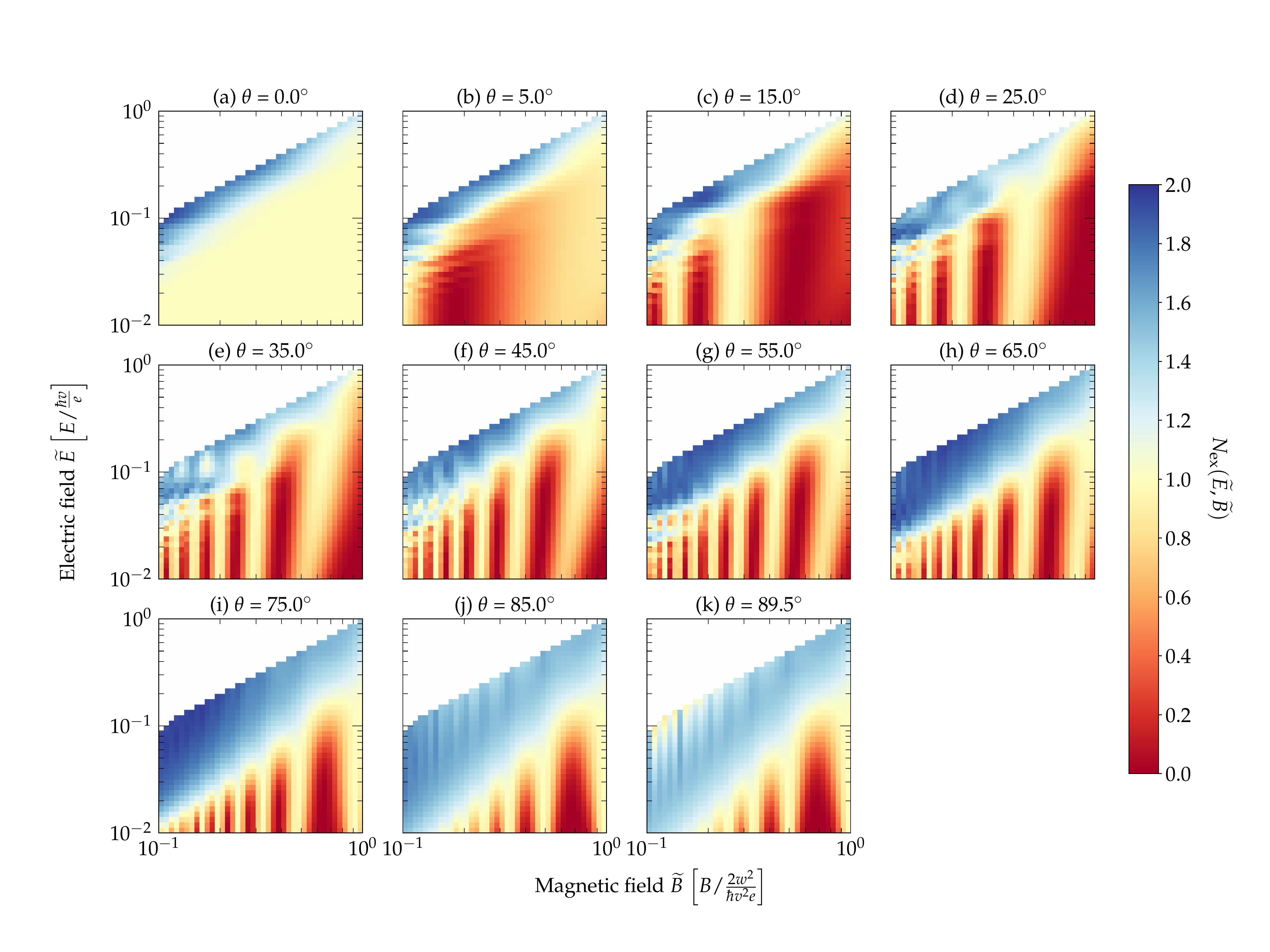}
	\caption{The expectation value $N_{\mathrm{ex}}(\widetilde{E},\widetilde{B})$ of the number of electrons in the conduction bands as a function of rescaled magnetic and electric fields - $\widetilde{B}$ and $\widetilde{E}$ in the region $\widetilde{B} < 1$.  All parameters of simulations are the same as for the phase diagrams presented in Fig.~\ref{fig:HSNL_phase_diagram}. (a) - (b) Disappearance of chiral-anomalous pumping if the direction of magnetic field without any additional structure. (c) - (k) Additional structures on disappearance of anomalous-chiral pumping due to gap closing of $\mathcal{H}_{\mathrm{NSNL}}$ at certain values of magnetic field. Note that this ladder structure shifts toward smaller $\widetilde{B}$ with increasing the deviation (the angle $\theta$) of applied magnetic field from the non-symmorphic nodal loop (NSNL) plane. \label{fig:HSNL_phase_diagram_details}}
\end{figure*}

In Fig.~\ref{fig:HSNL_phase_diagram}, we show the expectation value $N_{\mathrm{ex}}(\widetilde{E},\widetilde{B})$
as a function of the rescaled fields $\widetilde{E}$,$\widetilde{B}$ and the angle $\theta$ between the direction of the applied magnetic field and the plane of the NSNL in $k$-space.
The expected $N_{\mathrm{ex}}(\widetilde{E},\widetilde{B})$ for smaller values of $\widetilde{E}$ are presented in Fig.~\ref{fig:HSNL_phase_diagram_details}.

Results show that $N_{\mathrm{ex}}(\widetilde{E},\widetilde{B})$ is highly sensitive
to the direction of the applied magnetic field (the angle $\theta$) and pronounced for small values of $\widetilde{E}$ and $\widetilde{B}$
(i.e., weak electric and magnetic fields). The ladder of gap closing, see Fig.~\ref{fig:HSNL_Gaps},
manifests in $N_{\mathrm{ex}}(\widetilde{E},\widetilde{B})$, see Fig.~\ref{fig:HSNL_phase_diagram_details}. 

At $\theta = 90^{\circ}$ (the direction of the applied magnetic fields is perpendicular to the NSNL) the gap closes at $\widetilde{B} = 1/n$ (in chosen units),
for $n \in \mathbb{N}$. For $\theta < 90^{\circ}$ this ladder of gap closing is shifted towards weaker magnetic fields, see Fig.~\ref{fig:HSNL_Gaps}
and is observable only with sufficiently weak electric fields too, see Fig.~\ref{fig:HSNL_phase_diagram_details}. 

In the limit $\theta \to 0$, the ladder structure is completely suppressed by the presence of NSNL, see LLs in Fig.~\ref{fig:HSNL_LL_part_I} in Appendix~\ref{sec:spectra_HSNL}
and Fig.~\ref{fig:HSNL_Gaps}. In $N_{\mathrm{ex}}(\widetilde{E},\widetilde{B})$, the value 1 is exactly recovered which is changed only to higher values upon an application
of stronger electric fields. The critical value of electric field $\widetilde{E}$ to which this ladder of gap closing is observable is strongly dependent on the angle $\theta$ too,
see Fig.~\ref{fig:HSNL_phase_diagram_details}.

For a sufficiently low value of magnetic field $\widetilde{B}$ the gap is small and the NL is present, see the spectra in Appendix ~\ref{sec:spectra_HSNL}.
The length of the NSNL in the $k$-space is proportional to $\cos(\theta)$.
The length of the NL affects the maximal values of $\widetilde{E}$, see Fig.~\ref{fig:HSNL_phase_diagram_details}, where the non-trivial behavior is observable.

For the calculation of $N_{\mathrm{ex}}(\widetilde{E}, \widetilde{B})$, see Fig.~\ref{fig:HSNL_phase_diagram},
in each of four bands of $\mathcal{H}_{\mathrm{NSNL}}$ 5 LLs were considered leading to $N=9$ (after removing two ghost states) and Slater determinants with up to $M = 2$ excited electrons (implying overall 1378 Slater determinants). Overall 270 $k$-points were used in the simulation of real-time dynamics. The region approaching 
the di-adiabatic limit ($\widetilde{E} > \widetilde{B}$) was omitted since it cannot be studied reasonably with this value of $M$. Note that, for the proper identification of ghost states, the matrix representation of $\mathcal{H}_{\mathrm{NSNL}}$ is truncated at a much higher number (160) than just the number of considered LLs (18).

\section{Conclusions}
We demonstrated a simple numerical algorithm which allows for systematic inclusion of non-adiabatic contributions
to the quantum limit of chiral charge pumping (still close to the sufficiently weak electric fields). 
Considering Landau-Zener-like problems and non-interacting many-body $\bs{k}\cdot\bs{p}$ models,
we showed that, in these non-adiabatic corrections to the quantum limit of the chiral charge pumping,
the relative homotopy invariant~\cite{Sun2018} and Euler class invariant~\cite{Bouhon2020}
are non-trivially manifested. Moreover, this manifestation takes place before the onsets of 
annihilation of WPs with respect to the model free parameters, suggesting that if these non-adiabatic contributions
are measurable, they can act as an experimental probe for the mentioned topological invariants.
Furthermore, for non-symmorphic systems~\cite{Bzdusek2016}, we showed these contributions
to be highly sensitive to the direction of the applied magnetic field (with respect to the NSNL). 
It is possible that these contributions are measurable in 
longitudinal magnetoresistance, if they are not suppressed by other channels of conductivity in a particular material realization. 
The presented approach can be easily applied to other $\bs{k}\cdot\bs{p}$ (e.g., double Dirac semimetals~\cite{Wieder2016}) or tight-binding models.
It may also be interesting to examine, if there is a fingerprint of several nontrivial models of WSMs and Dirac semimetals
(e.g., models presented in Refs.~\cite{Sun2018,Bouhon2020,Bzdusek2016,Wieder2016}) in non-adiabatic contributions
to the $1/B$-quantum oscillations~\cite{Deng2019,Deng2019_2,Das2020} or to study the role of temperature through the density-matrix formalism.

\begin{acknowledgements}
M.B. acknowledges stimulating discussions with T.~Bzdušek and A.~Soluyanov. M.B. was supported by Comenius University under 
Grants for Young Researchers No. UK/436/2021 and No. UK/454/2022, and by the Slovak Academic Information Agency under the National Scholarship Programme of the Slovak Republic in 2018. 
Calculations were performed on Scoula Internazionale Superiore di Studi Avanzati Ulysses and ETH Euler clusters.
\end{acknowledgements}

\appendix

\section{Verification of the method on the single-node isotropic WP model \label{sec:demonstation_WSM}}

To verify the correctness of numerical calculations
[the value - $N_{\mathrm{ex}}(E,B)$],
we apply the method to a single-node (isotropic) WP model adapted from Ref.~\cite{Hosur2013}.
We do not aim to make any conclusion here about the non-adiabatic contributions
to the chiral charge anomaly with respect to the transport as a model of at least
two Weyl nodes of opposite chiralities would be needed; rather we want demonstrate
the correctness of the numerical results for a simple model when the Landau-Zener transition
can be inferred analytically.

We consider the ideal Weyl Hamiltonian as 
\begin{align}
    \mathcal{H} = \hbar v \bm{k}\cdot\bm{\sigma}\text{,}\label{eq:ideal_weyl}
\end{align}
where $\bm{\sigma} = (\sigma_x,\sigma_y,\sigma_z)$ are standard anticommuting $2\times2$ Pauli matrices, $v$ has the dimension of velocity, and $\bm{k}$ 
is the momentum. The energy spectrum of this Hamiltonian is $\eps_{\pm}(\bs{k})=\pm \hbar vk$. Considering the ideal Weyl Hamiltonian in parallel magnetic and electric fields, we can model the presence of a magnetic field by performing the Peierls substitution. We assume electric and magnetic fields to be
\begin{align}
    \bm{B} = (0,0,B)\, \quad\text{and}\quad  \,\bm{E}=(0,0,E).
\end{align}
The Peierls substitution changes the operators of momentum:
\begin{align}
    (\hbar k_x, \hbar k_y, \hbar k_z) &\mapsto (-i\hbar \partial_x + eA_x, -i\hbar\partial_y + eA_y, \hbar k_z)\nonumber \\
    &\equiv (\Pi_x,\Pi_y,\hbar k_z)\text{,}
\end{align}
and the canonical momentum commutator changes to
\begin{align}
    \left[\Pi_x,\Pi_y\right] = ie\hbar(\partial_yA_x-\partial_xA_y) = -ie \hbar B \text{.} 
\end{align}
Hence, the $\Pi$-operators can be expressed using the usual ladder operators $[\mathfrak{a},\mathfrak{a}^\dag]=1$ as
\begin{eqnarray}
    \Pi_x &=& \sqrt{\frac{e\hbar|B|}{2}}(\mathfrak{a}+\mathfrak{a}^\dag)\nonumber\\
    \Pi_y &=& i \mathrm{sign}(B)\sqrt{\frac{e\hbar|B|}{2}}(\mathfrak{a}-\mathfrak{a}^\dag)\text{.} 
\end{eqnarray}
Assuming $B>0$, the ideal Weyl Hamiltonian is changed into
\begin{align}
    \mathcal{H}(k_z) = v\begin{pmatrix} \hbar k_z & \sqrt{2e\hbar B} \mathfrak{a} \\  
              \sqrt{2e\hbar B}\mathfrak{a}^{\dag} & -\hbar k_z  
                        \end{pmatrix}\label{eq:weyl}\text{.} 
\end{align}
The corresponding energy spectrum is
\begin{align}
    \varepsilon_0(k_z) &= - v\hbar k_z\nonumber \\
    \forall \mathrm{n} \in \mathbb{Z} \backslash \{0\}: \varepsilon_{\mathrm{n}}(k_z) & = \mathrm{sign}(n)v\sqrt{\hbar^2k_z^2+2e\hbar B |n|}\text{.}    
\end{align}
We define rescaled fluxes of electric and magnetic fields and expressed energy in multiples of $\hbar v$,
\begin{align}
\widetilde{E} = eE/\hbar v,\,\,
\widetilde{B} = 2eB/\hbar,\,\,\textrm{and}\,\,
\widetilde{\eps} = \eps/(\hbar v)\,\text{.}
\end{align} 
 In this convention, $\mathcal{H}(k_z)$ takes the form:
\begin{align}
    \mathcal{\widetilde{H}}(k_z) &= \begin{pmatrix} k_z & \sqrt{\widetilde{B}} \mathfrak{a} \\  
              \sqrt{\widetilde{B}}\mathfrak{a}^{\dag} & - k_z  
                        \end{pmatrix}\text{,}\label{eq:dimensionless_Weyl}
\end{align}
and the TDSE takes form:
\begin{equation}
    i\partial_{k_z}\ket{\psi(k_z)} = -\frac{1}{\widetilde{E}}\mathcal{\widetilde{H}}(k_z)\ket{\psi(k_z)}\,\text{.}\label{eq:TDSE_WSM}
\end{equation}

\begin{figure}[t!]
	\includegraphics[width=8cm]{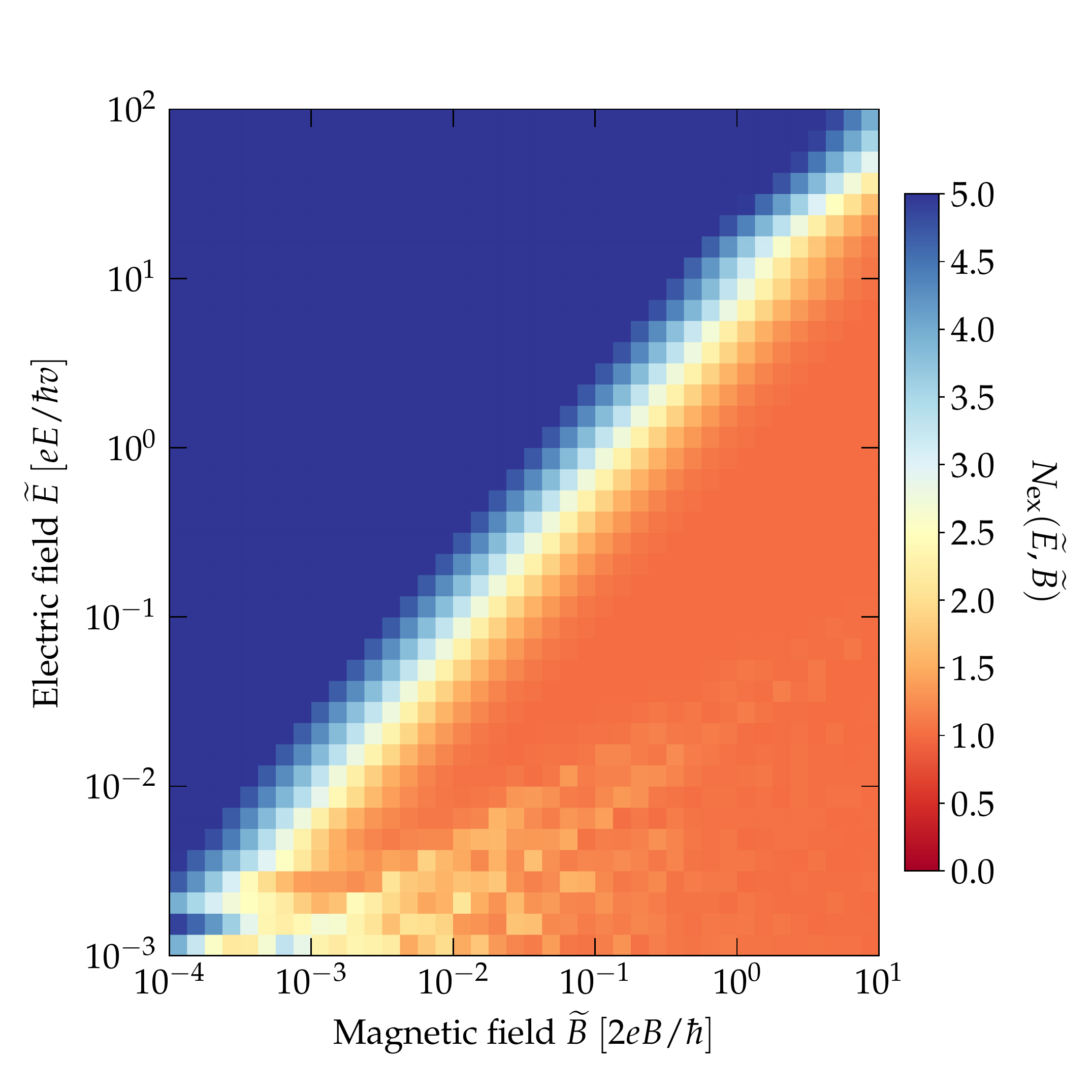}
	\caption{The expectation value $N_{\mathrm{ex}}(\widetilde{E},\widetilde{B})$ of number of electrons in the conduction bands as a function of rescaled magnetic and electric fields - $\widetilde{B}$ and $\widetilde{E}$ for ideal Weyl Hamiltonian given by Eq.~(\ref{eq:ideal_weyl}).  \label{fig:isotropic_WSM}}
\end{figure}

Results of our method for this simple model are presented in Fig.~\ref{fig:isotropic_WSM}. The diagram shows that for the Hamiltonian
given by Eq.~(\ref{eq:ideal_weyl}) the expectation value of $N_{\mathrm{ex}}(\widetilde{E},\widetilde{B})$
depends on $\widetilde{B}$ and $\widetilde{E}$ only via their ratio - $\widetilde{E}/\widetilde{B}$ as heuristically argued in Appendix~\ref{sec:argument}.
Note also a numerical noise visible in Fig.~\ref{fig:isotropic_WSM} in $N_{\mathrm{ex}}(\widetilde{E},\widetilde{B})$ in the region with $\widetilde{E} < 10^{-2}$
where the used step $\Delta k$ becomes inappropriate. The values of $\Delta k$ were kept constant in all calculations presented in the phase diagram.

\section{Heuristic argument for the dependence on the ratio - $\widetilde{B}/\widetilde{E}$ only in isotropic WSM \label{sec:argument}}
The single-electron TDSE given in Eq.~(\ref{eq:TDSE_WSM}) can be formally mapped to the class of Hamiltonians
studied by Brundobler and Elser~\cite{Brundobler1993} (to the generalization of the Landau-Zener problem). The corresponding 
TDSE in their notation becomes
\begin{align}
    \mathcal{H}(t) &= \mathcal{A} + \mathcal{B}t\,\text{,}\nonumber\\
    i\partial_t\psi(t) &= \mathcal{H}(t)\psi(t)\label{eq:BrundoblerTDSE}\,\text{,}
\end{align}
where $\mathcal{A}$ and $\mathcal{B}$ are constant Hermitian matrices. The identification is done as follows,
\begin{align}
    t &\to k_z\text{,}\nonumber\\
    \mathcal{A} &\to \begin{pmatrix} 0 & -\frac{\sqrt{\widetilde{B}}}{\widetilde{E}} \mathfrak{a} \\  
              -\frac{\sqrt{\widetilde{B}}}{\widetilde{E}}\mathfrak{a}^{\dag} & 0  
                        \end{pmatrix}\text{,}\nonumber\\
    \mathcal{B} &\to \begin{pmatrix} -\frac{1}{\widetilde{E}} & 0 \\  
              0 & \frac{1}{\widetilde{E}}
                        \end{pmatrix}\text{.}
\end{align}
Brundobler and Elser~\cite{Brundobler1993} found that elements of S-matrix for the time evolution given by Eq.~(\ref{eq:BrundoblerTDSE})
can be approximated and expressed in terms of the transition probabilities $p_{kl}$ between the two states 
[of $\mathcal{H}(t)$] $k$ and $l$ only,
\begin{equation}
    p_{kl} \approx \exp(-\pi z_{kl})\text{,}
\end{equation}
where $z_{kl}$ is the Landau-Zener parameter
\begin{equation}
    z_{kl} = \frac{{|\mathcal{A}_{kl}|}^2}{|\mathcal{B}_{kk} - \mathcal{B}_{ll}|} \sim \widetilde{B}/\widetilde{E}\text{.}
\end{equation}
Even though, we do not study the Landau-Zener problem of a single-electron Hamiltonian $\mathcal{H}(t)$, but rather a many-body Hamiltonian of
its tensor product, it gives insight, into why only the ratio of electric and magnetic fields (fluxes) should influence the non-adiabatic corrections 
to $N_{\mathrm{ex}}(\widetilde{E}, \widetilde{B})$, as demonstrated in Fig.~\ref{fig:isotropic_WSM}.

For the calculation of $N_{\mathrm{ex}}(\widetilde{E}, \widetilde{B})$, see Fig.~\ref{fig:isotropic_WSM},
overall 13 LLs (after removal of one ghost state) were considered ($N = 7$), with up to $M = 5$ excited electrons yielding to overall 1709 Slater determinants, and overall, 283 $k$-points were used for the real-time dynamics simulation.

\section{\label{sec:dot_product_Slater} Calculation of dot product of two Slater determinants in $\mathcal{O}(N^3)$}
Here we show how the dot product of two Slater determinants can be calculated in $\mathcal{O}(N^3)$ time. 
Considering two Slater determinants
\begin{align}
    \ket{\varphi_i} &= \frac{1}{\sqrt{N!}}\sum_\tau \mathrm{sign\,\tau}
    \left\{\tens{k} \ket{\psi_{\mathrm{\tau(\alpha_i(k))}}}\right\}\\ 
    \ket{\varphi^\prime_j} &= \frac{1}{\sqrt{N!}}\sum_{\tau^\prime} \mathrm{sign\,\tau^\prime}
    \left\{\tens{l} \ket{\psi^\prime_{\mathrm{\tau^\prime(\alpha_j(l))}}}\right\}\text{,}
\end{align}
we are interested into their dot product
\begin{align}
    &\braket{\varphi_i|\varphi^\prime_j} =\nonumber\\ 
    &\frac{1}{N!}\sum_{\tau^\prime}\sum_\tau\mathrm{sign\,\tau^\prime}\mathrm{sign\,\tau}\left[\prod_{k=1}^{N}\braket{{\psi_i}_{\tau(k)}|{\psi^\prime_j}_{\tau^\prime(k)}}\right]\text{.}
\end{align}

Note that although the direct evaluation of the dot product requires $(N!)^2$ operations, there are just $N^2$ 
independent terms. Note also that the determinant of the matrix $M_{ij} = \braket{\psi_i|\psi^\prime_j}$ can be expressed as
\begin{align}
    \det M \equiv \sum_{\tau^\prime} \mathrm{sign\,\tau\prime} \prod_{i=1}^N 
    \braket{\psi_i|\psi^\prime_{\tau(i)}}\,\text{,}
\end{align}
where $\tau'$ is the corresponding permutation.
Using $\det  (AB) =\det A \cdot \det B$, one can see that
matrix $M$ permuted by $\tau$ has a determinant $\det (\tau M)$ equal to 
$\mathrm{sign}\,\tau\,\det M$. 
On the other hand, $(\tau M)_{ij} = \braket{\psi_{\tau(i)}|\psi^\prime_j}$, and thus it follows that
\begin{align}
    \det (\tau M) &\equiv \sum_{\tau^\prime} \mathrm{sign\,\tau\prime} \prod_{i=1}^N 
    \braket{\psi_{\tau(i)}|\psi^\prime_{\tau(i)}}\nonumber\\
    &= \mathrm{sign}\,\tau\,\det M\text{.}
\end{align}
By multiplying both sides of the last equation by $\mathrm{sign}\,\tau$ and summing over all $N!$ permutations 
covered by $\tau$, we identify in $\det  M$ the dot product of two Slater determinants 
$\braket{\varphi_i|\varphi^\prime_j}$,
\begin{align}
    \braket{\varphi_i|\varphi^\prime_j} = \det M\text{.}
\end{align}
We note that this formula was spresented in~Ref.~\cite{Lowdin1955}.

\section{\label{sec:spectra_HSNL} LLs of model NSNL Hamiltonian}

\begin{figure*}[t!]
	\includegraphics[width=0.99\textwidth]{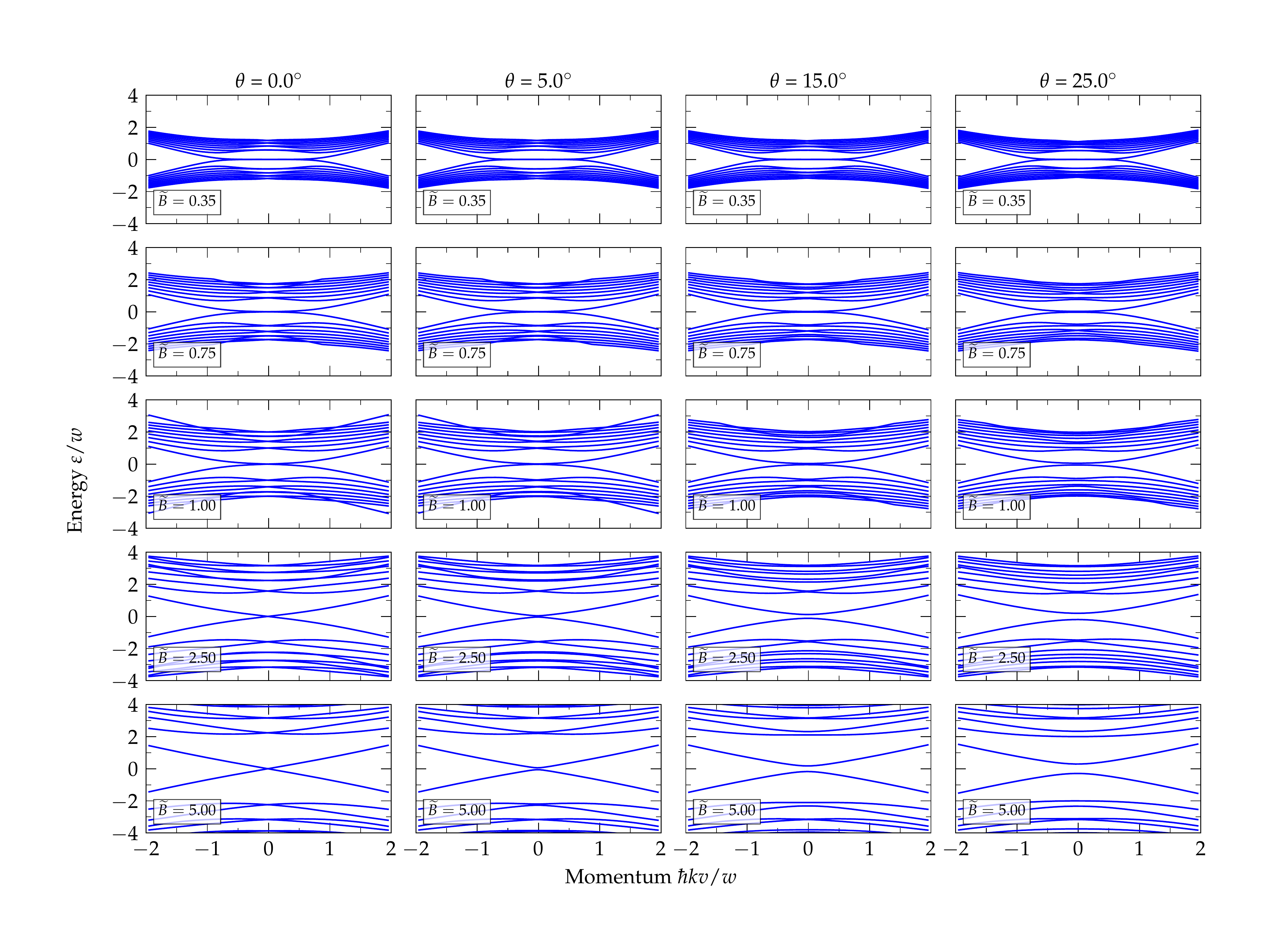}
	\caption{Landau levels (LLs) of $\mathcal{H}_\mathrm{NSNL}$. Nine LLs below and above the Fermi level are shown.\label{fig:HSNL_LL_part_I}}
\end{figure*}

\begin{figure*}[t!]
	\includegraphics[width=0.99\textwidth]{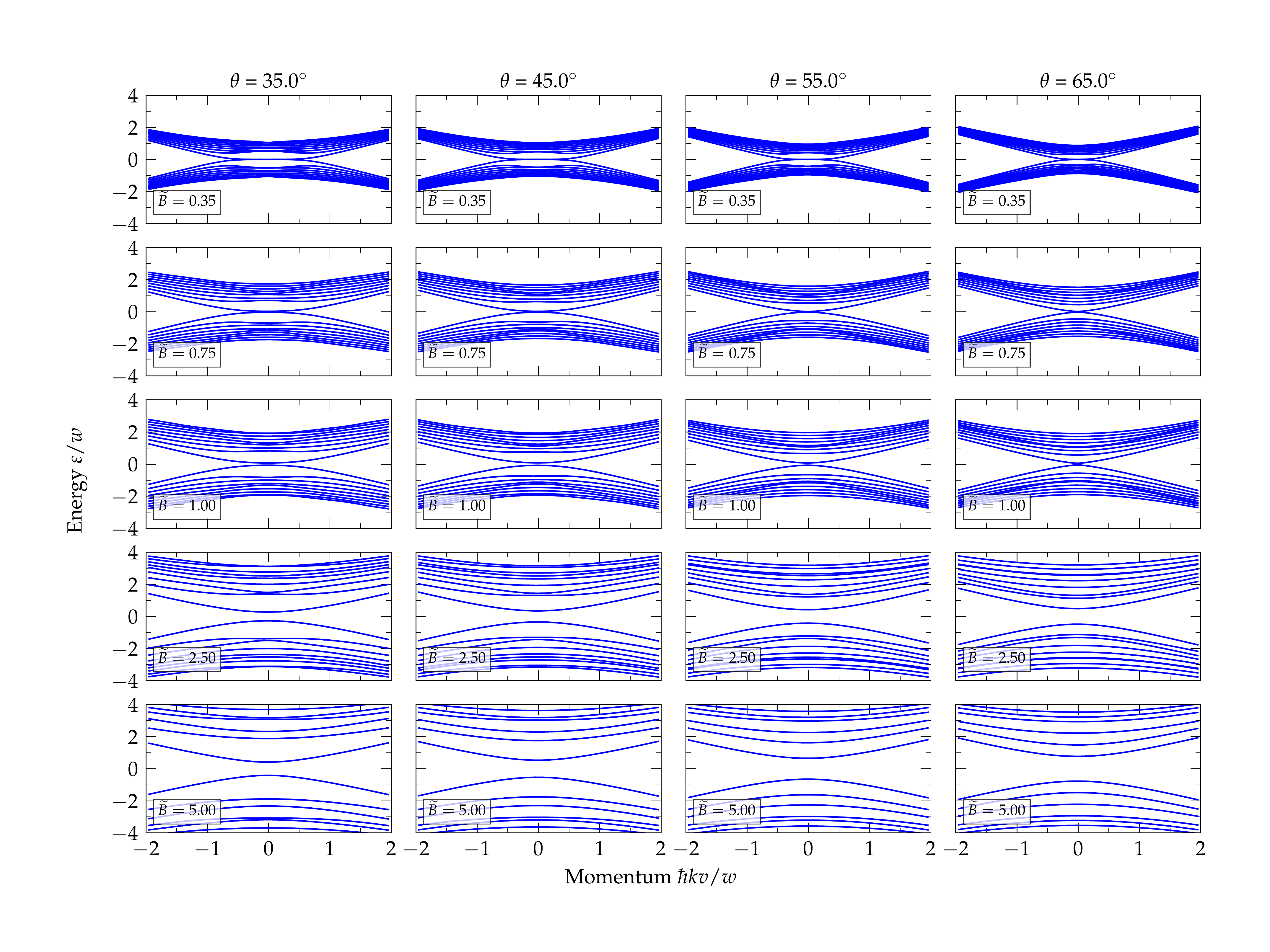}
	\caption{Landau levels (LLs) of $\mathcal{H}_\mathrm{NSNL}$. Nine LLs below and above the Fermi level are shown.\label{fig:HSNL_LL_part_II}}
\end{figure*}

\begin{figure*}[t!]
	\includegraphics[width=0.99\textwidth]{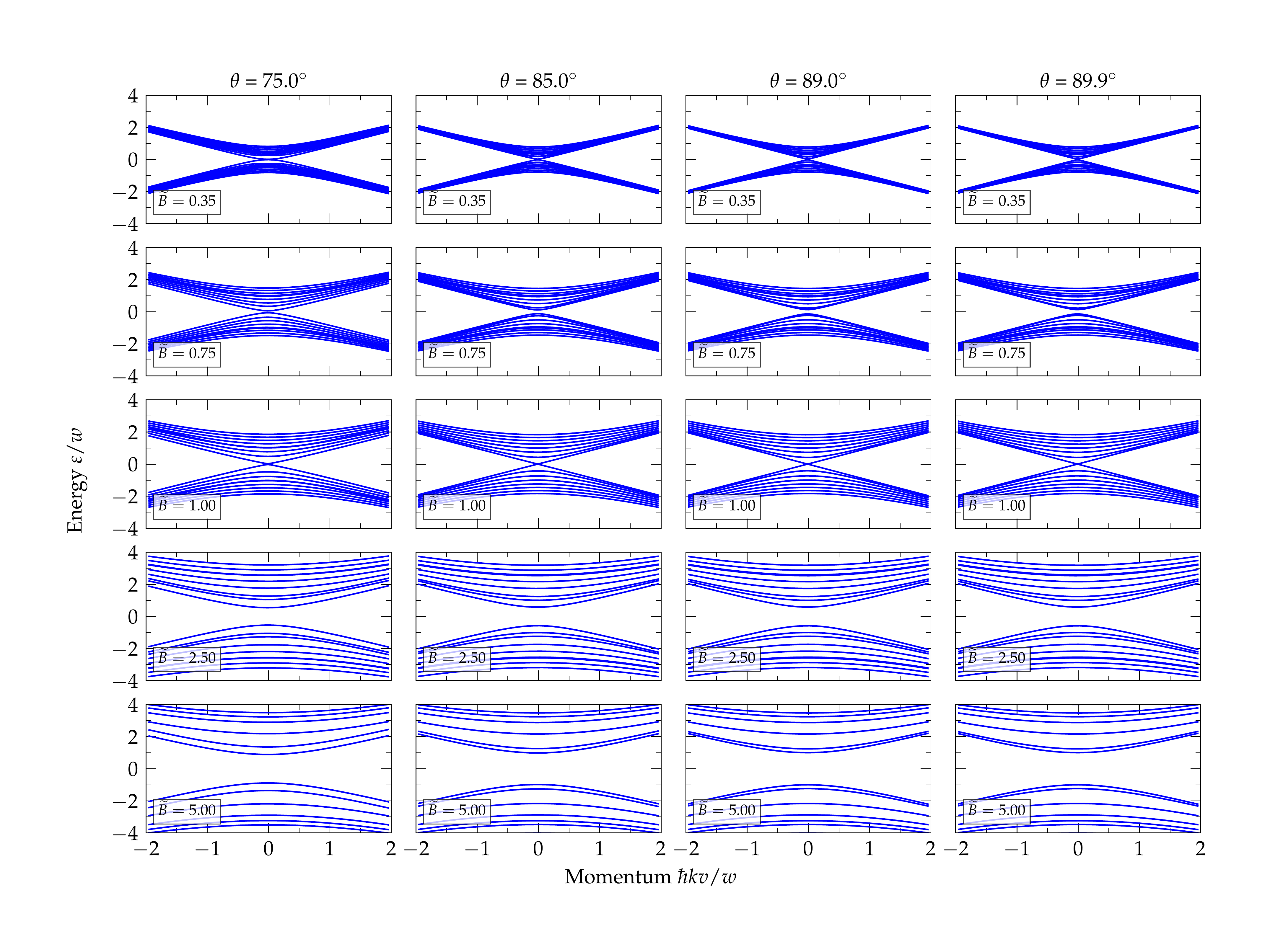}
	\caption{Landau levels (LLs) of $\mathcal{H}_\mathrm{NSNL}$. Nine LLs below and above the Fermi level are shown.\label{fig:HSNL_LL_part_III}}
\end{figure*}

In Figs.~\ref{fig:HSNL_LL_part_I} - \ref{fig:HSNL_LL_part_III} we show the LLs of the model NSNL Hamiltonian - $\mcH_\textrm{NSNL}$
given by Eq.~(\ref{eqn:NSNL-Ham}) as a function of angle $\theta$ between the direction of the applied magnetic field and the plane of the NSNL in the reciprocal space.

\end{document}